# Novel Mixed Approximate Deconvolution Subgrid-Scale Models for Large-Eddy Simulation

Ehsan Amani[1,*] (احسان امانی), Mohammad Bagher Molaei[1] (محمدباقر مولایی), and Morteza Ghorbani[2,3,4] (مرتضی قربانی)

[1]*Department of Mechanical Engineering, Amirkabir University of Technology (Tehran Polytechnic), Iran*
[2]*Faculty of Engineering and Natural Science, Sabanci University, 34956 Tuzla, Istanbul, Turkey*
[3]*Sabanci University Nanotechnology Research and Application Center, 34956 Tuzla, Istanbul, Turkey*
[4]*Faculty of Technology, Design and Environment, Oxford Brookes University, Headington, Oxford, OX3 0BP Wheatley, Oxford, OX33 1HX, UK*

**Abstract**

Approximate Deconvolution (AD) has emerged as a promising closure for Large-Eddy Simulation (LES) in complex multi-physics flows, where the conventional pure Dynamic Eddy-Viscosity (DEV) models experience issues. In this research, we propose novel improved mixed hard-deconvolution or secondary-regularization models and compare their performance with the existing standard mixed AD-DEV and penalty-term regularizations. For this aim, five consistency criteria, based on the properties of the modeled sub-filter-scale stress in limit conditions, are introduced for the first time. It is proved that the conventional hard-deconvolution models do not adhere to a couple of important primary criteria. Furthermore, through *a priori* and *a posteriori* analyses of Burgers turbulence and turbulent channel flow, it is manifested that the inconsistency with the primary criteria can result in larger modeling errors, the over-prediction and pile-up of kinetic energy in eddies of a length scale between the explicit filter width and grid size, and even the solution instability. On the other hand, the favorable characteristics of the new mixed models, in terms of the consistency criteria, significantly improve the accuracy of the predictions, the solution stability, and even the computational cost, particularly for one of the new models called mixed Alternative-DEV (A-DEV).

[*] Corresponding author. Address: Mechanical Engineering Dept., Amirkabir University of Technology (Tehran Polytechnic), 424 Hafez Avenue, Tehran, P.O.Box: 15875-4413, Iran. Tel: +98 21 64543404. Email: eamani@aut.ac.ir

## 1. Introduction

Large Eddy Simulation (LES) is recognized as a powerful tool for predicting complex turbulent flows in practical industrial applications. The accurate modeling of Sub-Grid Scale (SGS) stress, which governs the effect of the unresolved turbulent scales on the resolved ones, is the key factor to the success of the LES approach and has been an active area of research for decades.

Widely-used Eddy-Viscosity (EV) SGS models suffer from several drawbacks; Although the incorporation of backscatter is possible through a Dynamic EV (DEV) approach, the ensemble averaging which is usually performed to prevent the solution instability hinders the true backscattering (Maulik & San 2017). Furthermore, the validity of the underlying SGS model assumptions is questionable for multiphysics problems such as turbulent multiphase flows (Saeedipour *et al.* 2019). To address these issues, structural SGS closures could be employed. The precursor of the structural SGS approach is the scale similarity model proposed by Bardina *et al.* (1980). Layton and Lewandowski (2003) showed that the SGS stress modeled by this approach is reversible (Carati *et al.* 2001), accounts for backscatter, and best aligns with the true SGS stress tensor. However, the SGS dissipation of the scale similarity model is not sufficient. This model was then generalized to propose a family of structural models called Approximate Deconvolution (AD).

AD is one of the most intricate and promising SGS modeling strategies, originally proposed by Stolz and Adams (1999). This method is based on the explicit filtering of flow quantities without relying on any phenomenological assumptions. This approach has been conceptually borrowed from the image processing field (Biemond *et al.* 1990). Stolz and Adams (1999) and Stolz *et al.* (2001a) suggested that an effective deconvolution process is composed of two stages: the soft- and hard-deconvolution problems. The soft-deconvolution stage involves approximating the inverse filter through iterative methods, which allows for the recovery of the original scales up to the Nyquist wavenumber. The Van Cittert iterations (Germano 2009, Germano 2015) was adopted for this stage. However, the frequencies larger than the cut-off wavenumber are unrecoverable. This is because scales smaller than the grid length scale are lost, which leads to incomplete reconstruction of the actual quantities. To resolve this issue, a hard-deconvolution step, also known as the (secondary) regularization, is required to add the SGS effect on the deconvolved scales and drain the kinetic energy from the solution. If the regularization is not incorporated, energy can accumulate, leading to the instability of the solution in most cases. In the last two decades, the AD-LES technique has been employed to simulate a variety of problems of compressible and incompressible flows, including shock-boundary-layer interaction (Stolz *et al.* 2001b, a), atmospheric boundary layers (Chow *et al.* 2005), channel flows (Küng 2007), cavity flows (Habisreutinger *et al.* 2007), two-phase flows (Saeedipour *et al.* 2019), two-dimensional decaying homogeneous isotropic turbulence (Caban *et al.* 2022), compressible decaying isotropic turbulence of dense gases (Zhang *et al.* 2022), and turbulent jet flows with dilute spray droplets (Angelilli *et al.* 2022).

For successful implementation of AD-LES in a practical engineering problem, several major challenges, including the preservation of the solution stability and robustness, proper decomposition into the soft- and hard-deconvolution problems, accurate secondary regularization, efficient filter inversion algorithm (San & Vedula 2018), design of the explicit filter (Chang *et al.* 2022, Caban & Tyliszczak 2023), proper extension of the explicit filter to unstructured grids (Najafiyazdi *et al.* 2023) and parallel computing architectures (Kim *et al.* 2021), and efficient incorporation of low-dissipative numerical schemes on a general unstructured grid (Gärtner *et al.* 2020) and parallel computers (Boguslawski *et al.* 2024), have to be tackled. Considerable effort has been dedicated to providing solutions to these problems.

The type of the AD modeling strategy, i.e., the procedure of decomposition of Sub-Filter-Scale (SFS) stress into a deconvolved SFS stress, $b_{ij}$, and a modeled SFS stress, $a_{ij}$, is a key element of AD-LES. $b_{ij}$ is recoverable in terms of the deconvolved velocity (soft deconvolution), and $a_{ij}$ has to be modeled by an SGS closure (hard deconvolution). The most common AD types used in the literature are Simple ADM (SADM) (Stolz & Adams 1999), Scale-Similarity ADM (SSADM) (Stolz & Adams 1999), and ADM with a Relaxation Term (ADMRT) (Schlatter et al. 2004).

In addition to the type of AD, accurate hard deconvolution or (secondary) regularization is essential for ensuring the accuracy and robustness of the entire AD-LES and for preventing small-scale energy build-up. Originally, Stolz *et al.* (2001a) have presented a regularization method in which an *ad hoc* Penalty Term (PT) is added to the momentum equation. The model coefficient can be computed dynamically. For this purpose, two dynamic procedures have been proposed by Stolz *et al.* (2001a) and Sagaut (2006). However, both methods are purely dissipative and of high computational costs.

Alternatively, EV-based closures can be adopted for the hard-deconvolution step. Winckelmans and Jeanmart (2001) and Gullbrand and Chow (2003) combined the SADM AD-LES with the static Smagorinsky model and developed a static mixed model. However, this model is highly dissipative, and the issue of backscatter remains unresolved. In addition, the extension of the model to a dynamic variant is not straightforward (see section 2.2.3). Habisreutinger *et al.* (2007) adapted a mixed DEV closure to AD-LES with SSADM-type soft deconvolution, which was subsequently tested for cavity flow. They conducted a detailed analysis of the effects of SGS on AD-LES. Their results demonstrated a notable improvement resulting from the coupling of AD with the DEV model, and they successfully captured regions of backscatter. This mixed AD-LES approach was utilized in several subsequent studies, e.g., (Maulik & San 2018b, Saeedipour *et al.* 2019). Maulik and San (2017) evaluated several combinations of DEV and AD approaches for 2D isotropic decaying turbulence. Maulik and San (2018a) presented an alternative DEV approach, where the test filtering process for determining the model coefficient is replaced by the AD procedure. They tested their mixed model for the Burgers equation including shock discontinuities. Besides ADM, other relevant structural SGS closures have also shown improvements, in terms of stability and robustness, through hybridization with DEV. For instance, Yuan *et al.* (2021) used DEV to suppress instabilities in a class of deconvolution artificial neural network models, called dynamic iterative ADM; and Fan *et al.* (2023) utilized DEV to mitigate the stability issues of temporal direct deconvolution model, which is an ADM-like closure based on temporal rather than common spatial explicit filters (Oberle *et al.* 2024).

As introduced in this section, the accuracy of AD-LES has been widely recognized. However, the design of accurate as well as computationally efficient hard deconvolution is still a major challenge. Therefore, the main objective of this study is to propose improved mixed hard-deconvolution models, considering the significant influence of the hard deconvolution on the accuracy of AD-LES. Furthermore, the inconsistencies of the existing AD-LES models are explored and their performance is compared with the newly proposed models through extensive *a priori* and *a posteriori* analyses of three canonical Burgers turbulence tests and a turbulent channel flow benchmark.

The remaining sections of the paper are structured as follows. Section 2 presents the governing equations of AD-LES and decomposition to the soft- and hard-deconvolution steps. Moreover, several consistency criteria for the hard-deconvolution are proposed, the existing PT-based and mixed SGS closures are introduced, and the novel models are put forward. The detailed information on the benchmarks used to evaluate the models and numerical algorithms are reported in sections 3 and 4, respectively. Section 5 presents the results of the study, including the verification of the numerical solver as well as *a priori* and *a posteriori* analyses that carefully assess the characteristics of the models. Finally, in section 6, the main conclusions and most important findings of the study are summarized.

## 2. Mathematical modeling: Elements of AD-LES

### *2.1. AD-LES governing equations*

The Navier-Stokes equations for incompressible flows of a Newtonian fluid can be written as:

$$\partial_t u_i + \partial_j(u_i u_j) + \partial_i p - \partial_j(2\nu S_{ij}) \equiv \partial_t u_i + \mathcal{NS}(u_i) = 0 \qquad (1)$$

$$\partial_i u_i = 0 \qquad (2)$$

where $\partial_t$ and $\partial_i$ stand for the time derivative $(\partial/\partial t)$ and spatial derivative $(\partial/\partial x_i)$, respectively, and $u_i$ is the velocity vector, $p$ the (kinematic) pressure, $\nu$ the (kinematic) viscosity, and $S_{ij}$ the strain-rate tensor given as:

$$S_{ij} = \frac{1}{2}(\partial_j u_i + \partial_i u_j) \qquad (3)$$

A "kinematic" variable refers to the corresponding variable divided by the (constant) fluid density, $\rho$. Henceforth, the term "kinematic" is omitted and assumed implicit for all pressures, viscosities, and stresses. When solving equations on an LES grid, a grid filter, $\widetilde{(.)}$, is implicitly applied to Eqs. (1)-(3):

$$\partial_t \tilde{u}_i + \partial_j(\tilde{u}_i \tilde{u}_j) = -\partial_i \tilde{p} + \partial_j(2\nu \tilde{S}_{ij}) - \partial_j \tau_{ij} \qquad (4)$$

$$\partial_i \tilde{u}_i = 0 \qquad (5)$$

$$\tilde{S}_{ij} = \frac{1}{2}(\partial_j \tilde{u}_i + \partial_i \tilde{u}_j) \qquad (6)$$

where $\tau_{ij}$ is the SGS stress tensor defined by:

$$\tau_{ij} = \widetilde{u_i u_j} - \tilde{u}_i \tilde{u}_j \tag{7}$$

Apart from the modeling error, the numerical error of the discretization of spatial derivatives contributes to the total LES error. According to the modified wavenumber plot of the differentiation schemes (Stolz *et al.* 2002), the latter error is more pronounced for resolvable wavenumbers near the Nyquist frequency. Thus, it is desirable to suppress these wavenumbers in the solution, e.g., by employing an explicit filter with a non-dimensional cut-off wavenumber ($\omega'_C$) significantly smaller than the non-dimensional Nyquist wavenumber ($\omega'_N = \pi$). As a result, the numerical differentiation of this explicitly filtered solution is more accurate than the solution $\tilde{u}_i$, containing all wavenumbers up to the Nyquist frequency. Applying an explicit filter, denoted by $\overline{(.)}$ or $G * (.)$ where $G$ is the explicit filter kernel, to Eqs. (4) and (5):

$$\partial_t \bar{\tilde{u}}_i + \partial_j \left( \bar{\tilde{u}}_i \bar{\tilde{u}}_j \right) = -\partial_i \bar{\tilde{p}} + \partial_j \left( 2\nu \bar{\tilde{S}}_{ij} \right) - \partial_j T_{ij} \tag{8}$$

$$\partial_i \bar{\tilde{u}}_i = 0 \tag{9}$$

where $T_{ij}$ is the Sub-Filter-Scale (SFS) stress tensor defined by:

$$T_{ij} \equiv \overline{\widetilde{u_i u_j}} - \bar{\tilde{u}}_i \bar{\tilde{u}}_j = b_{ij} + a_{ij} \tag{10}$$

In AD-LES, Eqs. (8) and (9) are solved directly for $\bar{\tilde{u}}_i$; For this end, a closure is required for the unclosed SFS stress, $T_{ij}$, (or $\partial_j T_{ij}$). This quantity is decomposed into two parts, the deconvolved SFS stress, $b_{ij}$, and the modeled SFS stress, $a_{ij}$, based on the type of AD framework described in the next section. $b_{ij}$ can be expressed as a function of $\tilde{u}_i$ which is recoverable by a soft deconvolution while $a_{ij}$ has to be modeled by an SGS closure (hard deconvolution). Here, the Van Cittert iterative algorithm (Van Cittert 1931) is used to approximate $\tilde{u}_i$, i.e., an approximate deconvolution method. This method involves expanding the inverse explicit filter $G^{-1}$ as an infinite series, and then truncating it at the $N$-th term to obtain an approximation of $G^{-1}$, denoted by $Q_N$. The recursive form of the algorithm can be written as (Maulik & San 2018a):

$$u_0^\star = \bar{\tilde{u}}$$

$$u_n^\star = u_{n-1}^\star + (\bar{\tilde{u}} - G * u_{n-1}^\star) \ ; n = 1,2,\ldots,N \tag{11}$$

where $u^\star$ is the (deconvolved) approximation to $\tilde{u}$. Here, $N = 5$ is chosen.

For the decomposition of $T_{ij}$ into two contributions, $b_{ij}$ and $a_{ij}$, (soft-deconvolution models) and the closure for the modeled SFS stress, $a_{ij}$, (hard-deconvolution models), different approaches have been recommended in the literature. A decomposition of $T_{ij}$, used in the majority of previous works, see e.g., (Carati *et al.* 2001), is:

$$b_{ij} = \overline{\tilde{u}_i\tilde{u}_j} - \bar{\tilde{u}}_i\bar{\tilde{u}}_j \tag{12}$$

$$a_{ij} = \overline{\widetilde{u_i u_j}} - \overline{\tilde{u}_i\tilde{u}_j} = \bar{\tau}_{ij} \tag{13}$$

where $a_{ij}$ includes the SGS effect and has to be modeled ($a_{ij}^M$) by an SGS closure or (secondary) regularization, described in the next section. The superscript $M$ refers to the modeled counterpart of a quantity. Then, the deconvolved SFS stress can be written in the generalized SSADM form as (Stolz & Adams 1999):

$$b_{ij}^M = \overline{u_i^\star u_j^\star} - \overline{u_i^\star}\,\overline{u_j^\star} + \left(\overline{\tilde{u}_i\tilde{u}_j} - \overline{u_i^\star u_j^\star} + \overline{u_i^\star}\,\overline{u_j^\star} - \bar{\tilde{u}}_i\bar{\tilde{u}}_j\right) \tag{14}$$

The common approach is to neglect the terms in the parentheses on the right-hand side of Eq. (14), assuming $u_i^\star \approx \tilde{u}_i$, or more accurately absorbing them into $a_{ij}^M$.

Using the de-aliased formulation has been recommended by Carati *et al.* (2001) and Winckelmans and Jeanmart (2001), where the momentum equation (Eq. (8)) is recast and solved in the following form:

$$\partial_t\bar{\tilde{u}}_i + \partial_j\left(\widetilde{\bar{\tilde{u}}_i\bar{\tilde{u}}_j}\right) = -\partial_i\bar{\tilde{p}} + \partial_j\left(2\nu\bar{\tilde{S}}_{ij}\right) - \partial_j T_{ij}^D \tag{15}$$

$$T_{ij}^D \equiv \overline{\widetilde{u_i u_j}} - \widetilde{\bar{\tilde{u}}_i\bar{\tilde{u}}_j} = b_{ij}^D + a_{ij}^D = \tilde{T}_{ij} - \tilde{c}_{ij}; c_{ij} = \overline{\overline{u_i u_j}} - \overline{u_i u_j} \tag{16}$$

We call this approach "De-aliased ADM", adding prefix $D$ to its corresponding type acronyms and superscript $D$ to its corresponding variable symbols. For the de-aliased variants, $b_{ij}^D$ and $a_{ij}^D$ can be written in terms of $b_{ij}$ and $a_{ij}$ (of the regular AD) as:

$$b_{ij}^D = \tilde{b}_{ij} \tag{17}$$

$$a_{ij}^D = \tilde{a}_{ij} - \tilde{c}_{ij} \tag{18}$$

### *2.2. AD-LES SGS Models: The hard deconvolution*

Theoretically, all SGS closures can be adapted to model the effect of subgrid-scale motion in AD approaches, i.e., the term $a_{ij}$ in Eq. (13). In this section, first, we propose several consistency criteria in section 2.2.1 which can serve as guidelines to develop ADM closures. Then, the SGS models previously proposed for AD-LES in the literature are introduced in sections 2.2.2 and 2.2.3, and our new closure proposals are put forward in sections 2.2.4-2.2.6. The compatibility of each closure is also assessed versus the consistency criteria presented in section 2.2.1.

#### *2.2.1. Consistency criteria*

Assuming $\widetilde{(.)}$, $\overline{(.)}$, and $\widehat{(.)}$ indicate the grid filter, explicit filter, and test filter (in the mixed models), respectively. It is desirable that AD closures satisfy several consistency criteria:

- CI: In the limit of exact deconvolution ($N \to \infty$): $u_i^\star \to \tilde{u}_i$ while the closure for $a_{ij}$ should not approach a zero function, like its exact counterpart, Eq. (13), i.e.,:

$$u_i^\star \to \tilde{u}_i: a_{ij}^M \neq 0 \tag{19}$$

- CII: In the limit of all filter widths approaching zero ($\tilde{\Delta}, \bar{\Delta}, \hat{\Delta} \to 0$): $\hat{u}_i, \bar{u}_i, \tilde{u}_i \to u_i$, therefore, the closure for $T_{ij} = a_{ij} + b_{ij}$ should approach zero, like its exact counterpart, Eq. (10), and the DNS solution is retrieved, i.e.,:

$$\hat{u}_i, \bar{u}_i, \tilde{u}_i \to u_i: b_{ij}^M + a_{ij}^M = 0 \tag{20}$$

- CIII: In the limit of exact deconvolution ($N \to \infty$) and all filter widths approaching zero ($\tilde{\Delta}, \bar{\Delta}, \hat{\Delta} \to 0$): $u_i^\star \to \tilde{u}_i$ and $\hat{u}_i, \bar{u}_i, \tilde{u}_i \to u_i$, therefore, the closures for both $a_{ij}$ and $b_{ij}$ should approach zero, like their exact counterparts, Eqs. (12) and (13), i.e.,:

$$\hat{u}_i, \bar{u}_i, \tilde{u}_i, u_i^\star \to u_i: b_{ij}^M = a_{ij}^M = 0 \tag{21}$$

- CIV: If $A_{ij} = \mathcal{L}(a_{ij})$ for the exact modeled SFS tensors, $a_{ij}$ in Eq. (10) and $A_{ij}$ in Eq. (B2), where $\mathcal{L}$ is an arbitrary operator; the same should be true for their modeled counterparts:

$$A_{ij} = \mathcal{L}(a_{ij}): A_{ij}^M = \mathcal{L}(a_{ij}^M) \tag{22}$$

- CV: In the limit of exact deconvolution ($N \to \infty$) and only the grid size approaching zero ($\tilde{\Delta} \to 0$): $u_i^\star, \tilde{u}_i \to u_i$ and the closure for $a_{ij}$ should approach a zero function, like its exact counterpart, Eq. (13), i.e.,:

$$u_i^\star, \tilde{u}_i \to u_i: a_{ij}^M = 0 \tag{23}$$

### *2.2.2. Penalty Term (PT) regularizations*

Stolz *et al.* (2001a) suggested a penalty term through that $\partial_j a_{ij}$ part of $\partial_j T_{ij}$ in Eq. (8) is directly closed by:

$$\partial_j a_{ij}^M = \chi(I - Q_N * G) * \bar{\tilde{u}}_i = \chi(\bar{\tilde{u}}_i - \overline{u_i^\star}) \tag{24}$$

where $\chi$ is the dynamic model coefficient. Two methods for calculating $\chi$ have been introduced in the literature. Originally, Stolz *et al.* (2001a) presented the following equation, called Stolz and Adams Penalty Term (SAPT):

$$\chi(\boldsymbol{x}, t + \Delta t) = \chi_0 \frac{F_2(\boldsymbol{x}, t + \Delta t)|_{\chi=0} - F_2(\boldsymbol{x}, t)}{F_2(\boldsymbol{x}, t + \Delta t)|_{\chi=0} - F_2(\boldsymbol{x}, t + \Delta t)|_{\chi_0}} \tag{25}$$

where $F_2$ is the second-order structure function, and bold symbols are used to indicate the tensorial quantities. The discrete form of the second-order structure function in three-dimensional space at time $t$ and location $\boldsymbol{x}$ can be calculated by (Lesieur & Metais 1996):

$$F_2(\boldsymbol{x},t) = \langle \|\boldsymbol{\phi}(\boldsymbol{x}+\boldsymbol{r},t) - \boldsymbol{\phi}(\boldsymbol{x},t)\|^2 \rangle_{\|\boldsymbol{r}\|=h} \tag{26}$$

where $||\,.\,||$ and $\langle . \rangle_{\|\boldsymbol{r}\|}$ operators refer to the Euclidean norm of a vector and the local spatial averaging operator over a distance $\|\boldsymbol{r}\|$, respectively, $h$ the grid spacing, $\phi_i = \bar{\tilde{u}}_i - \overline{u_i^\star}$, and $\chi_0 = \chi(\vec{x},t)$ is the value of the dynamic coefficient at the beginning of a time step. For 3D uniform orthogonal structured grids, the operator $\langle . \rangle_{\|\boldsymbol{r}\|=h}$ in Eq. (26) can be simplified to a simple averaging over the six adjacent points while for unstructured grids, a weighted average over all neighboring cells is adopted (Lesieur & Metais 1996).

Sagaut (2006) proposed an alternative approach to calculate $\chi$, called Sagaut Penalty Term (SPT) here, as:

$$\chi = \frac{\left(\bar{\tilde{u}}_i - \overline{u_i^\star}\right)[G * \mathcal{NS}(u_i^\star)]}{\left(\bar{\tilde{u}}_i - \overline{u_i^\star}\right)\left(\bar{\tilde{u}}_i - \overline{u_i^\star}\right)} \tag{27}$$

where the operator $\mathcal{NS}(.)$ was defined in Eq. (1). Theoretically, this dynamic model was developed to enforce a constant kinetic energy level of the resolved sub-filter modes (Sagaut 2006), which are the scales between the explicit filter width and Nyquist.

The use of the de-aliased formulation by applying an extra grid-filter operation to the momentum advection term has been recommended, particularly for spectral solvers (Carati *et al.* 2001, Winckelmans & Jeanmart 2001). If the de-aliased form of the momentum equation, Eq. (15), is solved, e.g., D-SAPT, Eq. (24) is replaced with Eq. (28), including an extra grid filter:

$$\partial_j a_{ij}^{D,M} = \chi\left(\widetilde{\bar{\tilde{u}}_i - \overline{u_i^\star}}\right) \tag{28}$$

Now the consistency criteria introduced in section 2.2.1 are examined for the closures reported in this section. Based on Eq. (24), $a_{ij}^{M}$ approaches zero when $u_i^{\star} \rightarrow \tilde{u}_i$, and the present PT models are not consistent with CI. Likewise, when $\bar{u}_i, \tilde{u}_i \rightarrow u_i$, $a_{ij}^{M}$ does not tend to zero which is inconsistent with CII. On the other hand, it is straightforward to show that CIII is satisfied by Eq. (24).

### 2.2.3. *Standard Dynamic Eddy-Viscosity (S-DEV) closures*

EV SGS closures have also been used to account for SGS effect in AD-LES in the literature. Two different types of Dynamic EV (DEV) closures are considered here: The Dynamic Smagorinsky Model (DSM) (Germano *et al.* 1991) and Linear Dynamic Model with Equilibrium assumption (LDME) (Mokhtarpoor & Heinz 2017). The formulae associated with pure DEV-LES models are outlined in Appendix A.

A standard procedure has been used in the literature to use DEV model for the hard deconvolution in AD-LES, i.e., modeling $a_{ij}$ in Eq. (10). We refer to this standard procedure as Standard DEV (S-DEV) to distinguish it from the new strategies proposed in the present work, sections 2.2.4-2.2.6. In the standard framework, initially proposed by Winckelmans et al. (1998) for tensor-diffusivity models and then adapted to AD-LES by Habisreutinger *et al.* (2007), $a_{ij}$ in Eq. (10) is directly modeled by a DEV model as:

$$-a_{ij}^{M,r} = C_{\mathrm{D}} m_{ij}(\bar{\tilde{u}}_l);\ a_{ij}^{M,r} = a_{ij}^{M} - \frac{1}{3} a_{kk}^{M} \delta_{ij} \tag{29}$$

$$m_{ij}(\bar{\tilde{u}}_l) = 2\bar{\tilde{\Delta}}^2 \bar{\tilde{S}} \bar{\tilde{S}}_{ij} \tag{30}$$

where the superscript $r$ shows the deviatoric part of a tensor and $C_{\mathrm{D}}$ is the DEV coefficient. For the SSADM soft deconvolution, the model coefficient is calculated by (see Appendix B for the derivation):

$$C_{\mathrm{D}} = \frac{P^{r}_{ij} M_{ji}}{M_{mn} M_{nm}} \tag{31}$$

$$P_{ij} = \left(\widehat{\bar{\tilde{u}}_i \bar{\tilde{u}}_j} - \hat{\bar{\tilde{u}}}_i \hat{\bar{\tilde{u}}}_j\right) - \left(\widehat{\overline{u^{\star}_i}\, \overline{u^{\star}_j}} - \widehat{\overline{u^{\star}_i}}\, \widehat{\overline{u^{\star}_j}}\right) \tag{32}$$

To complete the modeling, $M_{ij}$ in Eq. (31) is defined based on the chosen DEV closure. Here, this term is expressed using the two closures introduced in Appendix A, i.e., Standard DSM (S-DSM) and Standard LDME (S-LDME) as:

$$M_{ij} = \widehat{m_{ij}(\bar{\tilde{u}}_l)} - m_{ij}\left(\hat{\bar{\tilde{u}}}_l\right); \quad \text{(DSM)} \tag{33}$$

$$M_{ij} = -m_{ij}\left(\hat{\bar{\tilde{u}}}_l\right); \quad \text{(LDME)} \tag{34}$$

The aforementioned S-DSM mixed AD-LES approach has been utilized in many previous works, e.g., (Habisreutinger *et al.* 2007, Maulik & San 2018b, Saeedipour *et al.* 2019).

If the de-aliased form of the momentum equation, Eq. (15), is solved, similar to the approach followed by Carati *et al.* (2001) and Winckelmans *et al.* (2001), it can be proved that $a^{D}_{ij}$ in Eq. (16) is modeled by Eqs. (29)-(34), replacing $a^{M}_{ij}$ with $a^{D,M}_{ij}$, $m_{ij}$ with $\tilde{m}_{ij}$, and $P_{ij}$ with $\tilde{P}_{ij}$.

Now the five consistency criteria introduced in section 2.2.1 are examined for S-DEV closure reported in this section. Based on Eq. (32), $P_{ij}$ (and $a^{M}_{ij}$) approaches zero when $\hat{u}_i, \bar{u}_i, \tilde{u}_i \to u_i$, and S-DEV is consistent with CII. Similarly, it is convenient to show that CIII- and CV-consistency are also maintained by S-DEV. In contrast, $P_{ij}$ (and $a^{M}_{ij}$) approaches zero when $u^{\star}_i \to \tilde{u}_i$, and S-DEV is not consistent with CI. In addition, from Eqs. (13) and (B8), it is inferred that $A_{ij} = \hat{a}_{ij}$; However, this relation does not hold between the modeled counterparts $a^{M}_{ij}$ and $A^{M}_{ij}$ defined by Eqs. (B4) and (B5). Therefore, S-DEV is inconsistent with CIV too. To remove the inconsistency issues of S-DEV, new proposals for the hard-deconvolution step are put forward in the subsequent sections.

### 2.2.4. *Modified Dynamic Eddy-Viscosity (M-DEV) closures*

In the first proposal, we suggest a Modified DEV (M-DEV) closure based on a slightly different decomposition of $T_{ij}^{T}$ defined in Eq. (B2). Instead of the decomposition given in Eq. (B8) for S-DEV, we propose:

$$B_{ij} = \widehat{\bar{\tilde{u}}_i \bar{\tilde{u}}_j} - \hat{\bar{\tilde{u}}}_i \hat{\bar{\tilde{u}}}_j, A_{ij} = \hat{T}_{ij}, B_{ij}^{M} = \widehat{\overline{u_i^{\star}}\,\overline{u_j^{\star}}} - \widehat{\overline{u_i^{\star}}}\,\widehat{\overline{u_j^{\star}}} \tag{35}$$

and keeping the decompositions in Eqs. (12) and (13) unchanged. These new decompositions for $B_{ij}$ and $A_{ij}$ are based on the fact that at the explicit filter level, i.e., Eqs. (8)-(10), $a_{ij}$ is equivalent to the explicit filter of SGS stress, or $\bar{\tau}_{ij}$, (see Eq. (13)). Therefore, at the test filter level, i.e., Eqs. (B1) and (B2), $A_{ij}$ can be assumed as the test filter of SFS stress, or $\hat{T}_{ij}$. Following, the least-squares procedure for the Leonard stress, described in Appendix B, the M-DEV closures for $a_{ij}^{M}$ are derived. These models are the same as Eqs. (29)-(34) with a difference that, instead of Eq. (32), $P_{ij}$ is given by:

$$P_{ij} = \widehat{\bar{\tilde{u}}_i \bar{\tilde{u}}_j} - \hat{\bar{\tilde{u}}}_i \hat{\bar{\tilde{u}}}_j - \left( 2\widehat{\overline{u_i^{\star}}\,\overline{u_j^{\star}}} - \widehat{\overline{u_i^{\star}}}\,\widehat{\overline{u_j^{\star}}} - \widehat{\overline{u_i^{\star} u_j^{\star}}} \right) \tag{36}$$

Again, the de-aliased variants of the models are applicable only by replacing $a_{ij}^{M}$ with $a_{ij}^{D,M}$, $m_{ij}$ with $\tilde{m}_{ij}$, and $P_{ij}$ with $\tilde{P}_{ij}$ in the present formulation. It is convenient to show that M-DEV is consistent with CI-CIV. However, M-DEV does not hold CV-consistency; it is anticipated that this inconsistency poses fewer issues than the other inconsistencies, provided that the ratio of $\bar{\Delta}/\tilde{\Delta}$ is not large, which is the case for the majority of LES computations. This point is examined in the results and discussion section.

### 2.2.5. *Enhanced Dynamic Eddy-Viscosity (E-DEV) closures*

In the second proposal, we suggest an Enhanced DEV (E-DEV) closure based on a slightly different interpretation of $T_{ij}^{T}$ defined in Eq. (B2). Comparing Eq. (B2) with Eq. (10), in the S-DEV approach (section 2.3.2), it was inferred that $T_{ij}^{T}$ resembles $T_{ij}$, replacing the filter $\overline{(.)}$ with the double-filter $\widehat{\widehat{(.)}}$. Alternatively, it is possible to conclude that $T_{ij}^{T}$ is obtained from $T_{ij}$ by replacing $\widetilde{(.)}$ with $\widehat{\widetilde{(.)}}$, knowing that the order of the filtering operations is interchangeable. As a result, instead of Eq. (B8), the decomposition of $T_{ij}^{T}$ becomes:

$$B_{ij} = \overline{\hat{\tilde{u}}_i \, \hat{\tilde{u}}_j} - \bar{\hat{\tilde{u}}}_i \bar{\hat{\tilde{u}}}_j, A_{ij} = \widehat{\widetilde{\overline{u_i u_j}}} - \overline{\hat{\tilde{u}}_i \hat{\tilde{u}}_j}, B_{ij}^{M} = \overline{\widehat{u_i^{\star}} \, \widehat{u_j^{\star}}} - \widehat{\overline{u_i^{\star}}} \, \widehat{\overline{u_j^{\star}}} \tag{37}$$

Following, the least-squares procedure for the Leonard stress, described in Appendix B, the E-DEV closures for $a_{ij}^{M}$ are derived. These models are the same as Eqs. (29)-(34) with a difference that, instead of Eq. (32), $P_{ij}$ is given by:

$$P_{ij} = \widehat{\bar{\tilde{u}}_i \bar{\tilde{u}}_j} - \bar{\hat{\tilde{u}}}_i \bar{\hat{\tilde{u}}}_j - \left( \overline{\widehat{u_i^{\star}} \, \widehat{u_j^{\star}}} - \widehat{\overline{u_i^{\star}}} \, \widehat{\overline{u_j^{\star}}} - \widehat{\overline{u_i^{\star} u_j^{\star}}} + \widehat{\overline{u_i^{\star}} \, \overline{u_j^{\star}}} \right) \tag{38}$$

Once again, the de-aliased variants of the models result simply by replacing $a_{ij}^{M}$ with $a_{ij}^{D,M}$, $m_{ij}$ with $\widetilde{m}_{ij}$, and $P_{ij}$ with $\tilde{P}_{ij}$ in the present formulation. It is convenient to show that E-DEV is consistent with all CI-CIV. CV-consistency is not obeyed by E-DEV similar to M-DEV.

### 2.2.6. *Alternative Dynamic Eddy-Viscosity (A-DEV) closures*

In the third group of novel models, referred to as Alternative DEV (A-DEV), in contrast to the former approaches described in sections 2.2.3-2.2.5, where $a_{ij}$ is directly modeled, a closure is sought for the SGS stress tensor, $\tau_{ij}$ in Eq. (7), and then the result is explicitly filtered to obtain $a_{ij}$ according to Eq. (13):

$$a_{ij}^{M,r} = \bar{\tau}_{ij}^{M,r} = G * \tau_{ij}^{M,r} \tag{39}$$

$$-\tau_{ij}^{M,r} = C_{\mathrm{D}} m_{ij}(\tilde{u}_l) = C_{\mathrm{D}} m_{ij}(u_l^{\star}) \tag{40}$$

$$m_{ij}(u_l^\star) = 2\tilde{\bar{\Delta}}^2 S^\star S_{ij}^\star \tag{41}$$

where the deconvolved velocity field, $u_i^\star$, is assumed as an approximation to $\tilde{u}_i$ overall this closure formulation. This process eliminates the need for extra explicit test filtering process, $\widehat{(.)}$ of width $\hat{\Delta}\sim 2\bar{\Delta}\sim 4\tilde{\Delta}$, which possesses a large width and involves the scales much larger than the grid filter in the DEV closure. To obtain the dynamic coefficient, $C_\mathrm{D}$, starting from Eq. (4) and applying a test box filter, $\breve{(.)}$, (as an approximation to the grid filter type) of width $\breve{\Delta}\sim \breve{\tilde{\Delta}} = 2\tilde{\Delta}$:

$$\partial_t \breve{\tilde{u}}_i + \partial_j\left(\breve{\tilde{u}}_i \breve{\tilde{u}}_j\right) = -\partial_i \breve{\tilde{p}} + \partial_j\left(2\nu \breve{\tilde{S}}_{ij}\right) - \partial_j \tau_{ij}^{T,r} \tag{42}$$

$$\tau_{ij}^T = \breve{\widetilde{u_i u_j}} - \breve{\tilde{u}}_i \breve{\tilde{u}}_j \tag{43}$$

Defining the Leonard stress, $L_{ij} = \tau_{ij}^T - \breve{\tau}_{ij}$, modeling $\tau_{ij}^T$ based on Eq. (40), i.e., $\tau_{ij}^{T,M} = C_\mathrm{D} m_{ij}\left(\breve{\tilde{u}}_l\right) = C_\mathrm{D} m_{ij}\left(\breve{u_l^\star}\right)$, and using the least-squares error minimization, the dynamic coefficient in Eq. (40) can be computed by :

$$C_\mathrm{D} = \frac{L_{ij}^r M_{ji}}{M_{mn} M_{nm}} \tag{44}$$

$$L_{ij} = \breve{\tilde{u}_i \tilde{u}_j} - \breve{\tilde{u}}_i \breve{\tilde{u}}_j \approx \breve{u_i^\star u_j^\star} - \breve{u_i^\star}\breve{u_j^\star} \tag{45}$$

where $M_{ij}$ in Eq. (44) is defined based on the chosen DEV closure. Here, this term is expressed using the two closures from Appendix A, i.e., A-DSM and A-LDME as (estimating $\tilde{u}_i$ with $u_i^\star$):

$$M_{ij} = \breve{m_{ij}(u_l^\star)} - m_{ij}\left(\breve{u_l^\star}\right);\ \text{(A-DSM)} \tag{46}$$

$$M_{ij} = -m_{ij}\left(\breve{u_l^\star}\right);\ \text{(A-LDME)} \tag{47}$$

It should be noted that in A-DEV, for modeling $b_{ij}$, the terms in the parentheses of Eq. (14) are omitted assuming $u_i^\star \approx \tilde{u}_i$ while in the previous closures, S-DEV, M-DEV, and E-DEV, these terms were accounted for by absorbing them into the directly modeled $a_{ij}$ tensor. The assumption of neglecting these terms for the A-DEV closures are assessed in the results and discussion section. The de-aliased variants of A-DEV models are applicable only by replacing $a_{ij}^{M}$ with $a_{ij}^{D,M}$, $\tau_{ij}^{M}$ with $\tau_{ij}^{D,M}$, $m_{ij}$ with $\widetilde{m}_{ij}$, and $L_{ij}$ with $\tilde{L}_{ij}$ in the present formulation. It is straightforward to show that A-DEV is consistent with CI-CIV. A-DEV does not adhere to CV-consistency; however, this raises issues if the ratio of $\breve{\Delta}/\tilde{\Delta}$ is large.

## 3. Benchmarks

### *3.1. The Burgers turbulence*

Burgers equation and Burgers turbulence are frequently utilized to test new numerical techniques (Edoh 2017) and turbulence models (Basu 2009, Heinz 2016, Li & Wang 2016, Bayona *et al.* 2018, Maulik & San 2018b, a, Amani *et al.* 2021, Subel *et al.* 2021), specifically those developed for resolving the Navier-Stokes equations. This is due to the similarity between Burgers and Navier-Stokes turbulence, including their advective nonlinear term balanced by diffusion, conservation of momentum and energy as viscosity approaches zero, separation of large and small scales with an inertial subrange, and intermittent energy dissipation (Frisch *et al.* 2001, Bec & Khanin 2007, Valageas 2009). However, it should be emphasized that the conclusions drawn from comparing closures for Burgers turbulence may not be directly applicable to the Navier-Stokes equations due to some nuanced dissimilarities between the two, such as the absence of scale invariance in Burgers turbulence. The general one-dimensional viscous forced Burgers equation is given by:

$$\frac{\partial u}{\partial t}+\frac{1}{2}\frac{\partial uu}{\partial x}=\nu\frac{\partial u^2}{\partial x^2}+f(x,t) \tag{48}$$

where $u$, $\nu$, and $f$ are the (dimensionless) velocity field, viscosity, and the forcing term, respectively. In the current study, the periodic boundary condition is used in the computational domain of length $2\pi$. The study also examines three different types of forcing through three different benchmarks described in Appendix C.

### 3.2. The turbulent channel flow

The turbulent flow of a Newtonian fluid between two parallel infinite plates, driven by a constant mean pressure gradient—referred to as "channel flow"—is a canonical problem extensively used for *a posteriori* and *a priori* analyses, owing to the availability of numerous Direct Numerical Simulation (DNS) databases. The dynamics of this problem is governed by a single dimensionless parameter, e.g., the friction Reynolds number, $Re_\tau = u_\tau H/\nu$, where $H$ represents the half channel height, $u_\tau = \tau_w^{1/2}$ the friction velocity, and $\tau_w$ the wall shear stress. In this investigation, the DNS dataset provided by Moser *et al.* (1999) is utilized. The computational domain is depicted in figure 1 with its dimensions as $(L \times 2H \times W) = (4\pi \times 2 \times 4\pi/3)$. The friction Reynolds number is considered to be $Re_\tau = 180$. The results are presented versus the normalized distance from the wall, $y^+ = yu_\tau/\nu$.

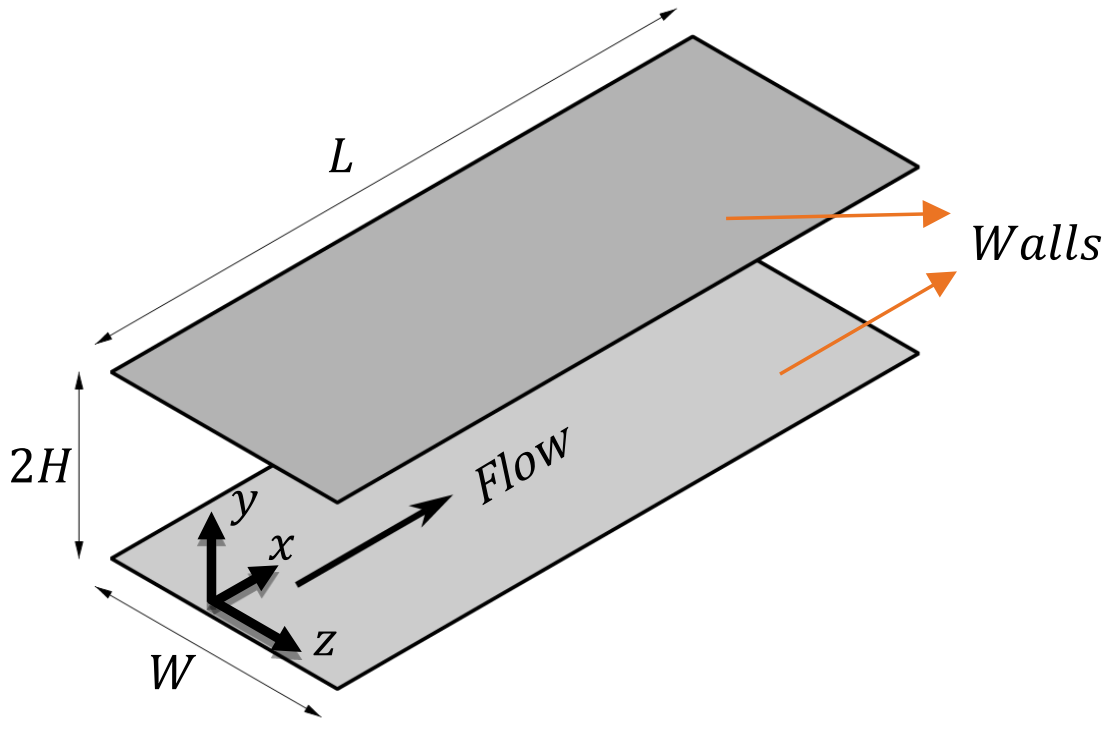

Figure 1: The schematic of the channel flow problem.

The flow statistics are reported based on the ensemble (Reynolds) averaging, e.g., the mean velocity, $\langle u_i \rangle$, Reynolds stress, $\langle u_i' u_j' \rangle$, etc.; where $u_i' = u_i - \langle u_i \rangle$ and $\langle \quad \rangle$ denotes the ensemble averaging operator, defined in the next section. These quantities are approximated in AD-LES, assuming $\langle q \rangle \approx \langle \bar{\tilde{q}} \rangle$ (Pope 2000), by:

$$\langle u_i \rangle \approx \langle \bar{\tilde{u}} \rangle; \ \langle S_{ij} \rangle \approx \langle \bar{\tilde{S}}_{ij} \rangle \tag{49}$$

$$\langle u_i' u_j' \rangle \approx \langle (\bar{\tilde{u}}_i - \langle \bar{\tilde{u}}_i \rangle)(\bar{\tilde{u}}_j - \langle \bar{\tilde{u}}_j \rangle) \rangle + \langle b_{ij} \rangle + \langle a_{ij}^r \rangle + \frac{2}{3} \langle \overline{k_{\mathrm{SGS}}} \rangle \delta_{ij} \tag{50}$$

where $k_{\mathrm{SGS}}$ is SGS Turbulent Kinetic Energy (TKE) and $a_{ij}$ is described based on the chosen SGS closures already introduced. For $k_{\mathrm{SGS}}$, the following estimation can be adopted (Sullivan et al. 2003):

$$k_{\mathrm{SGS}} \approx \left(\frac{C_{\mathrm{D}}}{C_{\mathrm{e}}}\right)^{\frac{2}{3}} \tilde{\Delta}^2 S^{\star 2} \;\; ; (\mathrm{A-DEV})$$

$$\overline{k_{\mathrm{SGS}}} \approx \left(\frac{C_{\mathrm{D}}}{C_{\mathrm{e}}}\right)^{\frac{2}{3}} \bar{\tilde{\Delta}}^2 \bar{\tilde{S}}^2 \; ; (\mathrm{S, M, E-DEV}) \tag{51}$$

where $C_e$ represents the model constant: $C_e = 1$ for LDME or $C_e = 1.048$ for DSM.

## 4. Numerical method

### *4.1. The Burgers turbulence benchmarks*

The DNS solutions of the Burgers equations, i.e., the solutions of Eq. (48), are obtained for the *a priori* analyses. For the *a priori* studies, the derivatives of the filtered properties are computed in the LES grid. The *a priori* tests are conducted using a box filter of width $\tilde{\Delta} = h_{\mathrm{LES}} = 16 h_{\mathrm{DNS}}$ for the injection or $= 64 h_{\mathrm{DNS}}$ for the other two benchmarks to obtain $\tilde{u}$ and a subsequent explicit filter to obtain $\bar{\tilde{u}}$, to be consistent with our *a posteriori* study.

For the *a posteriori* analyses, the explicitly filtered AD-LES equation corresponding to Eq. (48) is written as follows:

$$\frac{\partial \bar{\tilde{u}}}{\partial t} + \frac{1}{2}\frac{\partial \bar{\tilde{u}}\bar{\tilde{u}}}{\partial x} = \nu \frac{\partial \bar{\tilde{u}}^2}{\partial x^2} + \overline{\widetilde{f(x,t)}} - \frac{1}{2}\frac{\partial T}{\partial x} ; T = \overline{\widetilde{uu}} - \bar{\tilde{u}}\bar{\tilde{u}} \tag{52}$$

$$T = b + a = (\overline{\tilde{u}\tilde{u}} - \bar{\tilde{u}}\bar{\tilde{u}}) + (\overline{\widetilde{uu}} - \overline{\tilde{u}\tilde{u}}) \tag{53}$$

The explicit filter function (or convolution filter) is one of the most important factors influencing the performance of an AD-LES. For these test cases, we choose discrete compact

Padé filter function introduced by Lele (1992), successfully used in previous AD-LES studies, see, e.g., (Stolz & Adams 1999, Pruett & Adams 2000, Stolz *et al.* 2001a, San *et al.* 2015). For uniform structured grids, the one-parameter second-order implicit Padé filter, see, e.g., (San 2016), with $\alpha = 0.25$ for the filter width of $\bar{\Delta}= 2\tilde{\Delta}= 2h$ and with $\alpha = -0.25$ for $\hat{\Delta}= 2\bar{\Delta}= 4h$ is incorporated. The box filter is utilized as the test filter with twice the grid-size width, $\breve{\Delta}= 2\tilde{\Delta}= 2h$, in the A-DEV model and as the de-aliasing filter with the grid-size width, $\tilde{\Delta}= h$, in the de-aliased formulations. The discretized forms of the box filters on uniform structured grids can be found in reference (Davidson 2011). Spatial and temporal discretization has a significant effect on LES solution. Low-dissipative high-order discretization schemes are commonly used for AD-LES. Here, the fourth-order compact finite difference scheme (Lele 1992) for the spatial discretization and the fourth-order optimal Runge-Kutta scheme (Hirsch 2007) for the time derivatives is adopted. The verification of the numerical solvers is discussed in Appendix D.

### *4.2. The turbulent channel flow*

The design of explicit filters for unstructured grids is an open challenge in LES (Najafiyazdi *et al.* 2023). Kim *et al.* (2021) pointed out that recursive filters significantly reduce memory usage and enhance computational efficiency for domain-decomposed parallel LES. In this study, we use the recursive filtering approach, based on applying a narrow-band filter several times recursively to achieve filters with larger widths. Consequently, for the explicit filtering of width $\bar{\Delta}= 2\tilde{\Delta}= 2h$ or the box filter of width $\breve{\Delta}= 2\tilde{\Delta}= 2h$, the "simpleFilter" is incorporated which is defined by:

$$\bar{\phi}_i = \frac{\sum_{f\in\text{faces}} \phi_{f,i} |\mathbb{S}_{f,i}|}{\sum_{f\in\text{faces}} |\mathbb{S}_{f,i}|} \tag{54}$$

where $|\mathbb{S}_{f,i}|$ represents the area of face $f$ of grid cell $i$, $\phi_{f,i}$ is the interpolated value of the field $\phi$ on face $f$, and the summation is performed over all faces of cell $i$. This is an efficient filter that yields the whole filtered field together by an interpolation step from the cell centers to cell faces followed by a surface averaging step from the faces to centers. The transfer function of this filter is illustrated in figure 2. As it is observed, this filter possesses several necessary and favorable properties for explicit filtering, including normalization ($\hat{G}(\omega' = 0) = 1$), positiveness, smoothness, and vanishing at the Nyquist ($\hat{G}(\pi) = 0$), among others (Vasilyev *et al.* 1998, Najafiyazdi *et al.* 2023). Here, for the explicit filtering of width $\hat{\Delta}= 2\bar{\Delta}= 4h$, the "simpleFilter" is applied 5 times recursively. The number of filtering operations is chosen here based on the criterion of $\hat{G}(\pi/2) = 0.5$ or $\hat{G}(\pi/4) = 0.5$ (see figure 2) for the filter width of $2h$ or $4h$, respectively (Vasilyev *et al.* 1998, Stolz *et al.* 2001a).

Here, the cell-centered collocated Finite Volume Method (FVM) OpenFOAM CFD package, version 1912 (www.openfoam.com), was extended to include ADM-type closures. For the present channel flow LES, the pressure-velocity coupling is treated by the Pressure Implicit with Splitting of Operators (PISO) algorithm (Issa 1986) using two pressure-correction loops. For the discretization of time derivatives, the implicit second-order "backward" scheme is adopted (Greenshields 2015, Moukalled et al. 2016). Gradient calculations are performed using the Green-Gauss cell-based "Gauss linear" scheme (Greenshields 2015). The momentum advection term is discretized by the second-order pure-centered "Gauss linear" scheme, while the "linear" scheme is utilized for the interpolation of variables to cell faces. For the pressure equation, the system of discretized linear algebraic equations is solved using the Preconditioned Conjugate Gradient (PCG) method, complemented by the Diagonal Incomplete-Cholesky with Gauss-Seidel (DICGaussSeidel) preconditioner (Saad 2003). For the momentum, the Stabilized Preconditioned Bi-Conjugate Gradient (PBiCGStab) solver, with the Diagonal Incomplete-LU (DILU) preconditioning, is employed (Van der Vorst 1992, Barrett *et al.* 1994, Saad 2003). The convergence criterion is set to a normalized residual tolerance of $10^{-6}$ for all variables at each time step. To ensure stability and accuracy, the time step is dynamically adjusted to maintain the maximum Courant-Friedrichs-Lewy (CFL) number of 0.5.

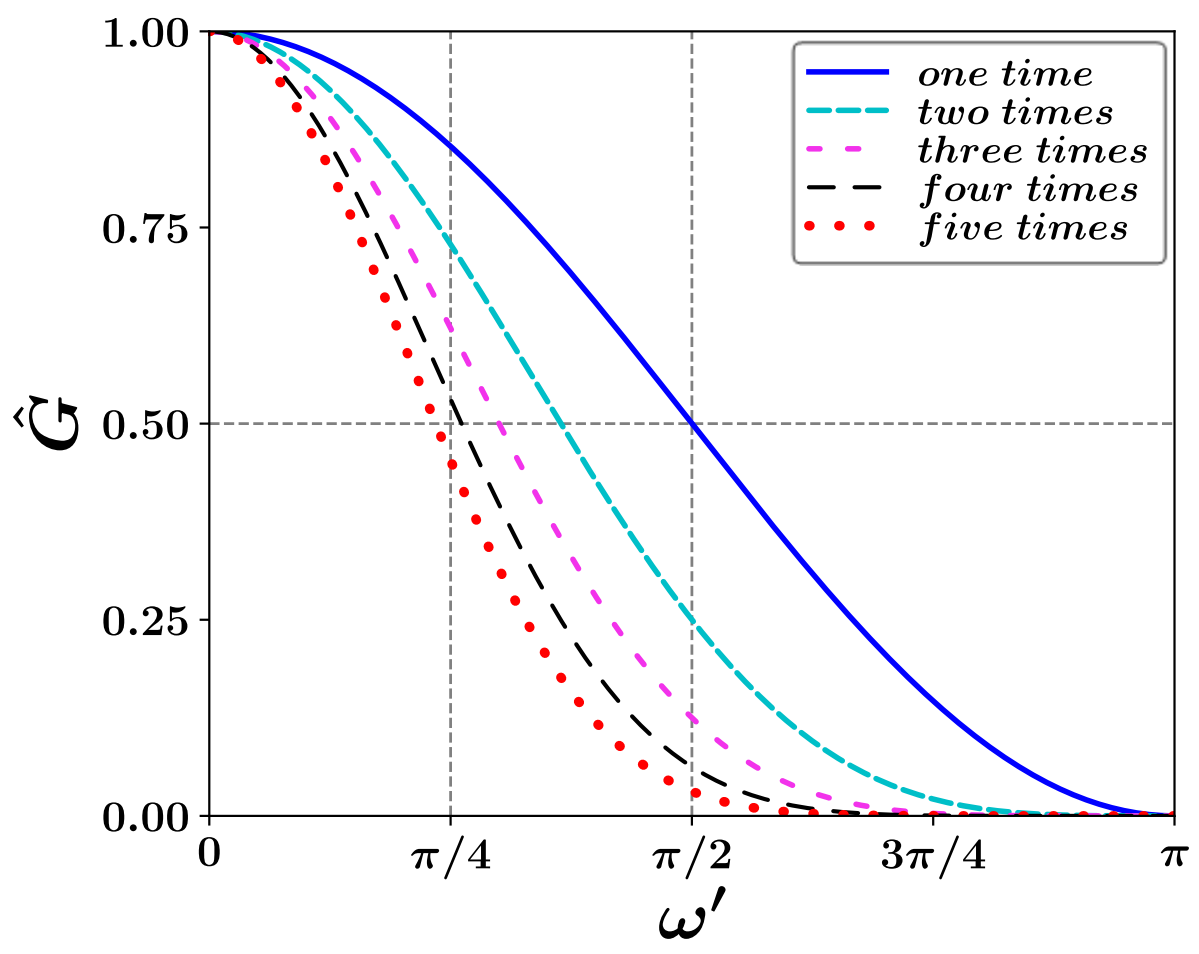

Figure 2: The recursive filter transfer function for different numbers of filtering operation.

The computational domain is illustrated in figure 1. Periodic boundary conditions are enforced in the streamwise and spanwise directions for all flow variables, enforcing the mean pressure gradient by an extra source term in the momentum equation (Amani *et al.* 2023). At the lower and upper walls, the no-slip boundary condition is applied to the velocity and a zero-gradient boundary condition to the pressure. A grid with uniform distribution in the streamwise and spanwise directions is devised; while in the wall-normal direction, a refinement towards the walls is applied using a hyperbolic tangent function with a target wall-normal distance of $y_w^+ = 0.82$. A slightly coarser grid is also generated to check the sensitivity of the models to the grid resolution. The specification of the present LES grids and reference DNS one is reported in table 1.

Table 1: Comparison of the present LES and reference DNS (Moser et al. 1999) grids.

| | $\Delta x^+$ | $\Delta y^+$ | $\Delta z^+$ | $\widetilde{\Delta}/h_{DNS}$ | $N_x \times N_y \times N_z$ |
|---|---|---|---|---|---|
| **DNS** | 17.7 | [0.06, 4.4] | 5.9 | 1 | $\mathbf{128 \times 129 \times 128 = 2113536}$ |
| **LES (default)** | 38.9 | [0.82, 22.08] | 14.78 | 4.15 | $\mathbf{58 \times 54 \times 51 = 159732}$ |
| **LES (coarse)** | 61.23 | [0.78, 29.08] | 20.98 | 5.43 | $\mathbf{37 \times 45 \times 36 = 59940}$ |

To accelerate the simulations, the Wall-Adapting Local Eddy-viscosity (WALE) model (Nicoud & Ducros 1999) is utilized for generating an initial solution for the other models, due to its robustness and simplicity. The velocity field initialization procedure and the solution steps to achieve a WALE model solution have been detailed elsewhere (Amani *et al.* 2023, Taghvaei & Amani 2023). Starting from the WALE solution, the simulations continue for 30 $FTT$, where $FTT$ is the Flow-Through-Time ($FTT = L/U_b$) and $U_b$ is the channel bulk velocity, marking the attainment of a statistically stationary condition. Subsequently, the simulations continue to collect the statistics by the ensemble averaging, $\langle \quad \rangle$, taken here as a time-averaging for 70 $FTT$ and spatial averaging in the two homogeneous directions ($x$ and $z$). For comments on the numerical solver validation, please see Appendix D.

### *4.3. General numerical considerations*

For the stability of the penalty-term regularizations, SAPT and SPT, introduced in section 2.2.2, several important points have to be taken into consideration. Firstly, the dynamic coefficient, $\chi$, approaches infinity when $\bar{\tilde{u}} \approx \overline{u^{\star}}$, which occurs in the case of insufficient turbulence or a laminar initial field. In such cases, the modeled SFS tensor, $a$, is not comparable in magnitude to the other terms, such as advection. To address this issue, we disregard the SFS tensor computation when its contribution is not on par with the other terms. Therefore, as the first guess, it is assumed that $\chi = 1/dt$, then the absolute value of the SFS tensor is calculated and compared with the absolute value of the convection term. If the ratio is less than 0.01, the calculation of $\chi$ and SFS term are omitted. Secondly, to mitigate the instability problems, the computed dynamic coefficient, $\chi$, over the domain is smoothed by applying a sufficiently wide filter (Padé filter with $\alpha = -0.25$) as recommended by Stolz *et al.* (2001a) on the structured grid and a test filter of width $4h$ on the unstructured grid. Thirdly, in order to maintain numerical stability during time advancement, the upper limit of $\chi$ is bounded by the inverse of the time-step size, while the lower limit is set at zero, $0 \leq \chi \leq 1 / dt$ (Stolz et al. 2001a). Lastly, for SAPT at the first time step of simulation, $\chi_0$ is initialized by $1 / (2dt)$ and at least two to four loops are necessary to achieve sufficient convergence for $\chi_0$. We found this critical to prevent instability issues in the simulations. At subsequent time steps, PT models require additional loops to update $\chi$; SAPT requires two extra loops for $F_2(\boldsymbol{x}, t + \Delta t)|_{\chi=0}$ and $F_2(\boldsymbol{x}, t + \Delta t)|_{\chi_0}$, while SPT requires an extra loop for $\mathcal{NS}(u_i^\star)$. Therefore, PT-based models involve computationally intensive algorithms compared to the other hard-deconvolution closures.

Diverging solutions were observed in conventional pure DEV and mixed AD-DEV models. As a potential solution, the clipping method based on the realizability conditions proposed by Mokhtarpoor and Heinz (2017) was implemented, but it was not sufficient to resolve the instability issue for many cases. Subsequently, for all pure DEV and mixed AD-DEV models, the Positive Total Viscosity (PTV) approach was employed as $\nu_t + \nu \geq 0$ or $\nu_t/\nu \geq -1$, unless stated otherwise; which successfully addressed the stability concerns.

## 5. Results and discussion

This section focuses on evaluating the performance and accuracy of the novel hard-deconvolution procedures for AD-LES and their comparison with the existing alternatives. Specifically, we examine two types of hard deconvolutions: PT-based and mixed AD-DEV. The dynamic coefficient in PT-based methods is calculated via two approaches: SAPT and SPT. Additionally, we assess four groups of mixed models, namely the conventional S-DEV and newly proposed M-DEV, E-DEV, and A-DEV, along with the LDME and DSM DEV closures. Furthermore, we investigate the effect of utilizing de-aliased forms of these models. The naming convention for the models is summarized in table 2.

Table 2: AD-LES hard-deconvolution models. The extra prefix D- refers to the usage of the de-aliased momentum equation, Eq. (15), instead of the default form, Eq. (8). DEV stands for Dynamic Eddy-Viscosity closure: LDME or DSM. Assuming the SSADM soft deconvolution, $b_{ij}^{M}$ is modeled using Eq. (14) (dropping the terms in the parentheses).

| **Hard deconvolution** | **$a_{ij}^{M}$ formulation** | **Consistency Criterion** CI Eq. (19) | CII Eq. (20) | CIII Eq. (21) | CIV Eq. (22) | CV Eq. (23) | **Comments** |
|---|---|---|---|---|---|---|---|
| PT | Eq. (24) | × | × | ✓ | ✓ | ✓ | - PT=SAPT: Eq. (25) → $\chi$<br>- PT=SPT: Eq. (27) → $\chi$<br>- The D- variant† |
| S-DEV | Eqs. (29)-(32) | × | ✓ | ✓ | × | ✓ | - DEV=DSM: Eq. (34) → $M_{ij}$<br>- DEV=LDME: Eq. (33) → $M_{ij}$<br>- The D- variant‡ |
| M-DEV | Eqs. (29)-(31), (36) | ✓ | ✓ | ✓ | ✓ | × | |
| E-DEV | Eqs. (29)-(31), (38) | ✓ | ✓ | ✓ | ✓ | × | |
| A-DEV | Eqs. (39)-(41), (44), (45) | ✓ | ✓ | ✓ | ✓ | × | - DEV=DSM: Eq. (47) → $M_{ij}$<br>- DEV=LDME: Eq. (46) → $M_{ij}$<br>- The D- variant* |

† $b_{ij}^{D,M} = \tilde{b}_{ij}^{M}$ and $\partial_j a_{ij}^{D,M} = \widetilde{\partial_j a_{ij}^{M}}$

‡ $b_{ij}^{D,M} = \tilde{b}_{ij}^{M}$ and $a_{ij}^{D,M}$ has the same formulae as $a_{ij}^{M}$, replacing $m_{ij}$ with $\tilde{m}_{ij}$ and $P_{ij}$ with $\tilde{P}_{ij}$

* $b_{ij}^{D,M} = \tilde{b}_{ij}^{M}$ and $a_{ij}^{D,M}$ has the same formulae as $a_{ij}^{M}$, replacing $m_{ij}$ with $\tilde{m}_{ij}$ and $L_{ij}$ with $\tilde{L}_{ij}$

The original AD-LES formulation was put forward using the SAPT regularization (Stolz *et al.* 2001a). Moreover, in the supplementary material, two dynamic PT regularizations, i.e., SAPT and SPT, have been compared for the Burgers turbulence tests, and it has been shown that SAPT yields more accurate predictions. Therefore, SAPT is chosen as the reference hard deconvolution in the subsequent sections against which other closures are compared.

### *5.1. Mixed AD-DEV closures: Burgers turbulence*

The PT hard deconvolutions are computationally expensive due to the procedure required for determining their dynamic coefficient, $\chi$. It involves several additional loops on the entire set of continuity and momentum equations to achieve sufficient convergence. The primary objective of this section is to assess the accuracy of less computationally-intensive DEV SGS models when used in conjunction with ADM and evaluate the performance of the novel closures proposed in this research. In figure 3, the comparison of two standard mixed AD-DEV closures, namely S-DSM and S-LDME (see also table 2), with the PT-based SAPT model is provided. This comparison demonstrates that the standard coupling of the SSADM approach with the DEV SGS models produces results comparable to SAPT; However, stronger oscillations, spreading far from the shock region (figure 3a), are observed in the velocity field, and a larger overshoot is noticed near the Nyquist wavenumber, which is manifested as a peak at the end of the energy spectrum (figure 3b). These oscillations were not evident in any of the PT-based closures (see the supplementary material). These fluctuations may be associated with limitations in the eddy viscosity models for the modeled SFS stress and/or additional aliasing errors.

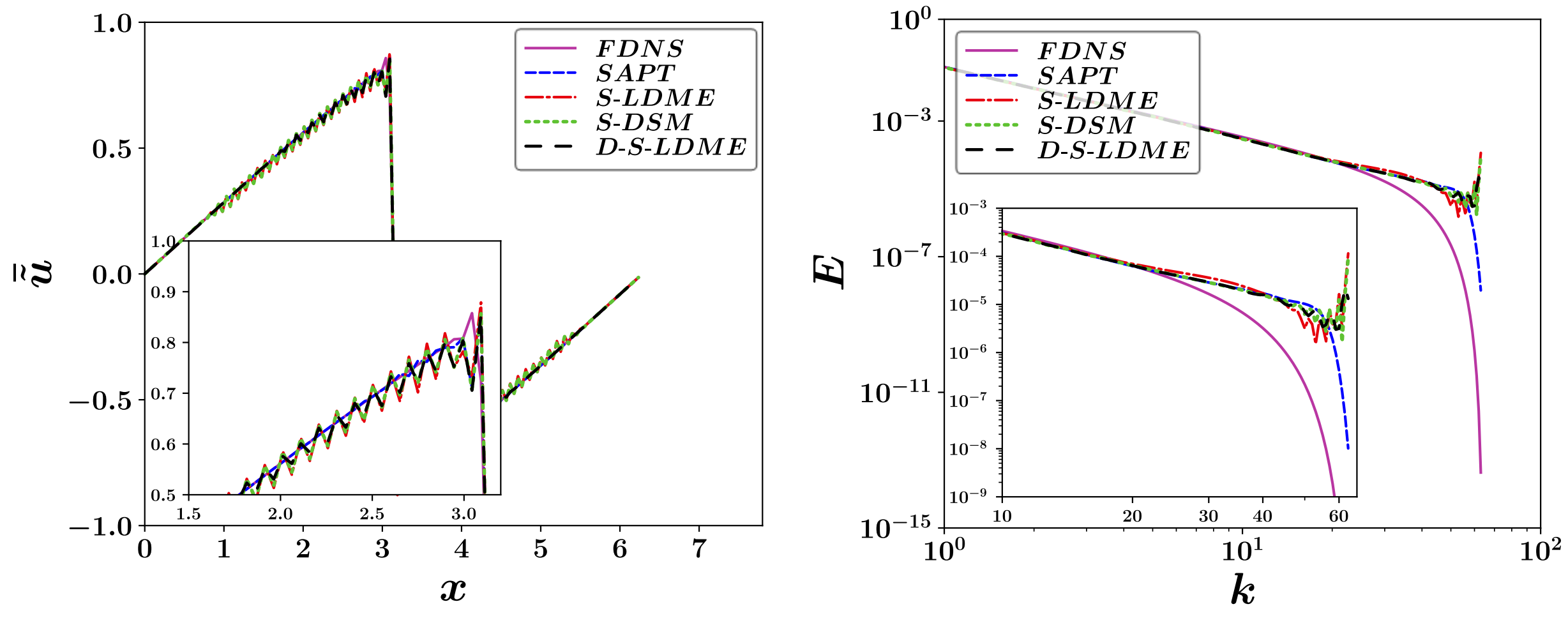

Figure 3: The "decaying" benchmark: a) The filtered velocity field and b) its energy spectrum, with two standard mixed AD-DEV closures and the PT-based SAPT against the FDNS result at $t = 2.5$.

To discover the main cause of the observed unphysical fluctuations, the study examines the prediction by a pure-DEV SGS closure, namely pure-LDME. Figure 4 depicts the velocity field and energy spectrum resulting from this model, and the findings suggest that the strong fluctuations are a characteristic feature of eddy-viscosity models. In the EV closures, the SGS stress is modeled as a non-linear function of the strain rate, $\tilde{S}_{ij}$, see Eqs. (A1) and (A2). The extra spatial differentiation of the filtered velocity field required to calculate $\tilde{S}_{ij}$, Eq. (6), induces additional errors particularly near the shock region where there exists considerable numerical dispersion error in the solution, leading to the aforementioned intensified fluctuations. Two reasons could explain the reduced oscillations in the standard mixed S-LDME compared to pure-LDME. Firstly, the utilization of the explicitly filtered strain rate, $\bar{\tilde{S}}_{ij}$, instead of $\tilde{S}_{ij}$ helps attenuate fluctuations. This is evidenced by the comparison of $\bar{\tilde{u}}$ and $\tilde{u}$ fields in figure 4a. The mitigated level of unphysical oscillations in $\bar{\tilde{u}}$ compared to $\tilde{u}$ reduces the uncertainty in the calculation of $\bar{\tilde{S}}_{ij}$ compared to $\tilde{S}_{ij}$. Secondly, in the mixed models, only a portion of the total SFS stress is modeled by the EV-based closure. For instance, based on the *a priori* analysis of the decaying benchmark, only 60% of SFS is modeled (through $a$), while the other 40% is deconvolved in terms of $b$. Note that these contributions are reported based on the ratio $\mathcal{N}(da/dx)/\mathcal{N}(dT/dx)$~0.6 from the *a priori* analysis where the norm operator, $\mathcal{N}(.)$, is defined by:

$$\mathcal{N}(Q) = \langle Q^2 \rangle_{L,\mathcal{T}}^{1/2} \tag{55}$$

and $\langle . \rangle_{L,\mathcal{T}}$ indicates the averaging over all domain, $L$, for the decaying case and the domain as well as the period of sampling, $\mathcal{T}$, for the two other cases. In addition, based on the *a priori* analysis, the contribution of the terms in the parentheses of Eq. (14) is less than 1.5%, justifying the omission of these terms in the A-DSM approach pointed out in section 2.2.6.

The elevated high-frequency noise, represented as an overshoot close to the Nyquist wavenumber in the spectrum of the mixed model (see S-LDME in figure 3b) as well as the spectrum of the pure EV (figure 4b), is associated with the extra aliasing error induced by the highly non-linear closure of DEVs. This error can be mitigated by resorting to a de-aliased formulation as can be advocated by comparing S-LDME with its de-aliased variant, i.e., D-S-LDME, in figure 4b. The de-aliased variant effectively decreases the ending overshoot. Therefore, to avoid the interference of the aliasing error with the other errors stemming from the model inconsistencies and uncertainties, the de-aliased forms of models are compared and analyzed, hereafter.

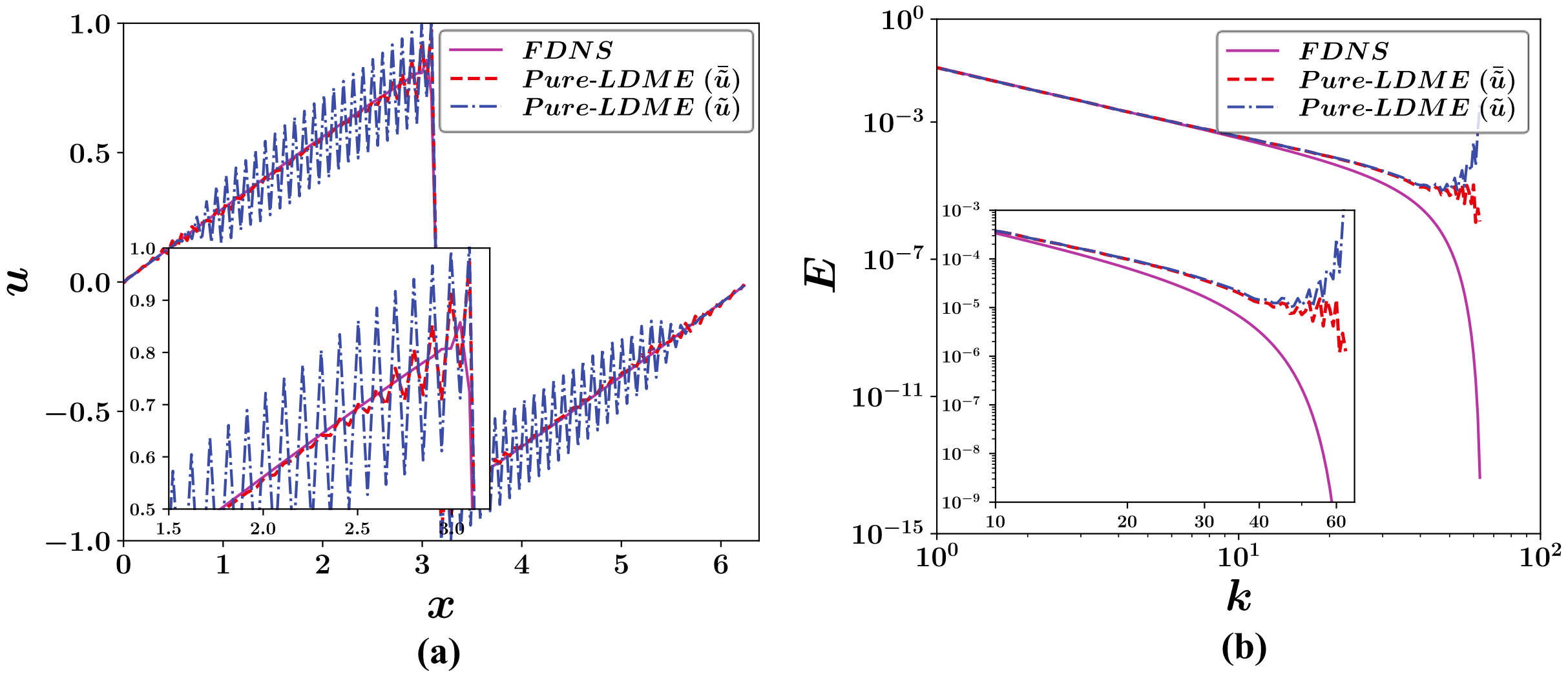

Figure 4: The “decaying” benchmark: a) The filtered velocity field and b) its energy spectrum, with the pure eddy-viscosity closure LDME at $t = 2.5$. Both $\bar{\tilde{u}}$ and $\tilde{u}$ fields have been plotted.

To assess the performance of the new mixed closures, namely M-DEV, E-DEV, and A-DEV (see also table 2), in comparison to the standard mixed S-DEV or PT-based model, the predictions using these models (de-aliased forms) are illustrated against FDNS in figure 5. It is interesting to see that all our new proposals, i.e., M-LDME, E-LDME, and A-LDME, exhibit higher accuracies than the standard S-LDME for the decaying benchmark. Specifically, E-LDME and A-LDME demonstrate great performances in attenuating the fluctuations of the velocity field and conforming to the FDNS energy spectrum.

To further investigate the reasons behind this improved performance, *a priori* tests are conducted on these models and the computed modeled stress, $a$, by these approaches is plotted in figure 6. According to this data, E-LDME prediction is much closer to the exact modeled stress. On the other hand, it becomes evident that the primary reason for the poor performance of the standard model compared to those of new models is its inability to accurately model the modeled SFS stress near the shock. This issue originates from inconsistency CI in S-LDME (see table 2 and section 2.2.1). This is because, in the *a priori* analysis, the situation dictates $u_i^\star \to \tilde{u}_i$ since the SGS is fully resolved. Meanwhile, $P_{ij}$ approaches zero for S-LDME according to Eq. (32), leading to a zero modeled SFS stress, $a_{ij}^M$, as can be seen in figure 6. This is not consistent with the exact definition of $a_{ij}$ (Eq. (13)), also plotted in figure 6. All new proposals are CI-consistent (please, see table 2) and this is the main reason underlying their improved performance.

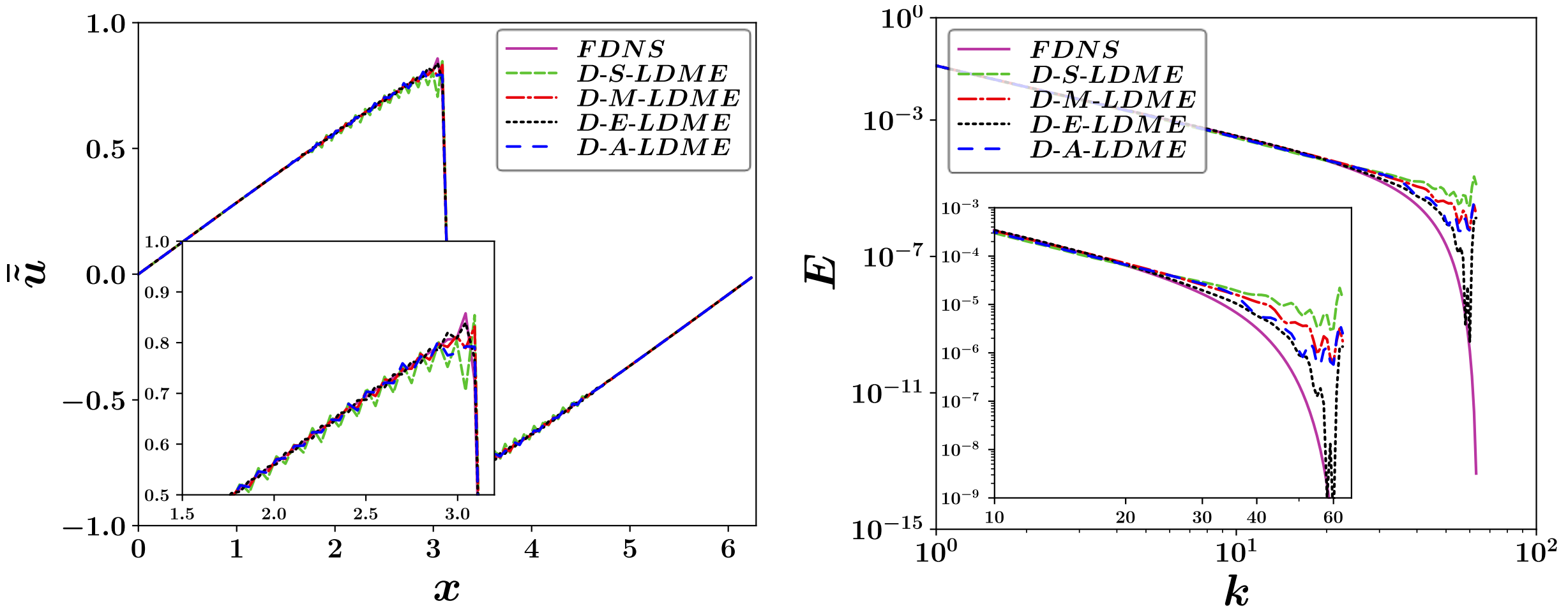

**(a)** **(b)**

Figure 5: The “decaying” benchmark: a) The filtered velocity field and b) energy spectrum, with different mixed AD-DEV approaches against FDNS at $t = 2.5$.

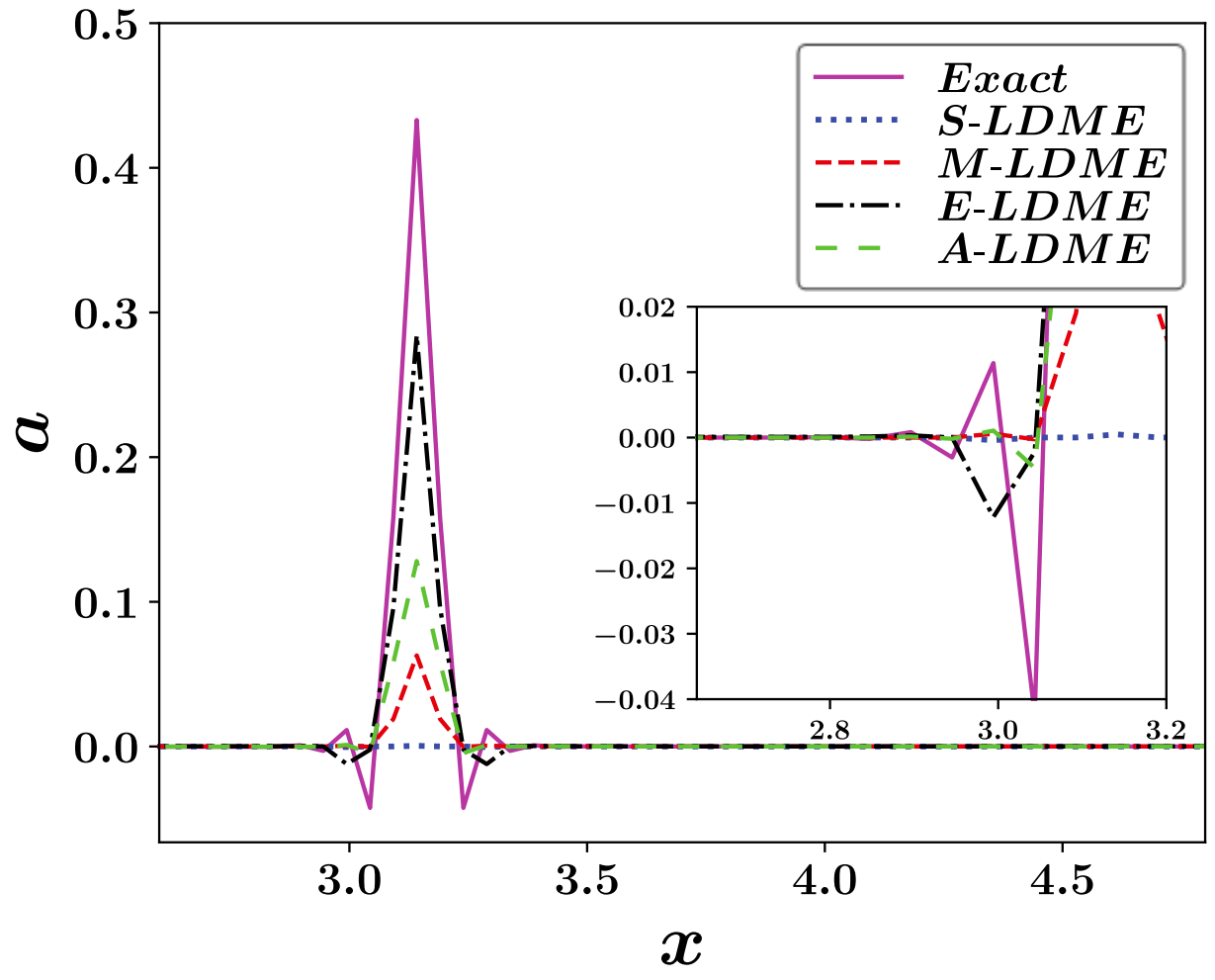

Figure 6: The *a priori* test of the “decaying” benchmark: The modeled SFS stress distribution at $t = 2.5$. The comparison of the exact form and mixed AD-DEV closures.

Next, the performance of the new models is further evaluated by *a priori* analysis. Table 3 presents the correlation coefficients of the modeled, $a$, deconvolved, $b$, and total, $T = a + b$, SFS stresses for the new and standard mixed approaches. Since all the AD-LES in this table use SSADM as the soft deconvolution, the correlation coefficients of $b$ are the same for all mixed models. On the other hand, the correlation coefficient of $a$ for all new models are around 0.99 which is 7% greater than the standard mixed model, S-LDME. The highest correlation coefficient in the *a priori* test of the decaying case belongs to A-LDME. However, E-LDME showed the highest accuracy in the *a posteriori* test (figure 5). This suggests that the correlation coefficient may not be an accurate measure of the accuracy at least for this benchmark. To check this point, the norm of error of $a$, $b$, and $T$ are also computed by the *a priori* test and reported in table 4. The error norm of each quantity $Q$ is computed by:

$$EN(Q) = \frac{\mathcal{N}(Q_{\text{LES}} - Q_{\text{FDNS}})}{\mathcal{N}(Q_{\text{FDNS}})} \tag{56}$$

According to table 4, the smallest error of $a$ as well as of $T$ pertains to the E-LDME closure followed by A-LDME and M-LDME, and the S-LDME has the largest error. This is entirely in line with the conclusions drawn from the *a posteriori* analysis. Therefore, the error norm may be a more pertinent measure of the accuracy than the correlation coefficient in *a priori* analyses.

Table 3: The *a priori* test of the "decaying" benchmark: The correlation coefficients of the modeled, $a$, deconvolved, $b$, and total, $T = a + b$, SFS stresses for the mixed standard, S-LDME, and new, M-LDME, E-LDME, and A-LDME, approaches.

| | **S-LDME** | **M-LDME** | **E-LDME** | **A-LDME** |
|---|---|---|---|---|
| $a$ | 0.923 | 0.988 | 0.989 | 0.992 |
| $b$ | 0.999 | 0.999 | 0.999 | 0.999 |
| $T = a + b$ | 0.966 | 0.981 | 0.995 | 0.996 |

Table 4: The *a priori* test of the "decaying" benchmark: The error norm of the modeled, $a$, deconvolved, $b$, and total, $T = a + b$, SFS stresses for the mixed standard, S-LDME, and new, M-LDME, E-LDME, and A-LDME, approaches.

| | **S-LDME** | **M-LDME** | **E-LDME** | **A-LDME** |
|---|---|---|---|---|
| $a$ | 0.94 | 0.90 | 0.57 | 0.81 |
| $b$ | 0.036 | 0.036 | 0.036 | 0.036 |
| $T = a + b$ | 0.61 | 0.58 | 0.37 | 0.52 |

Lastly, as another comparison between the mixed closures, the Probability Density Function (PDF) of the velocity field for each approach is presented in figure 7 against the FDNS. For moderate velocities, E-LDME and M-LDME demonstrate the best accuracy. In the vicinity of the shock region, where the velocity exhibits the largest positive or negative values, the performance of the models can be assessed through the level of alignment of the PDF tails with FDNS. As can be observed in figure 7, the E-LDME tail is closer to FDNS than the one of M-LDME. The results of the *a priori* and *a posteriori* tests advocate the significant improvement achieved by the new mixed models compared to the conventional S-DEV closure for the decaying benchmark.

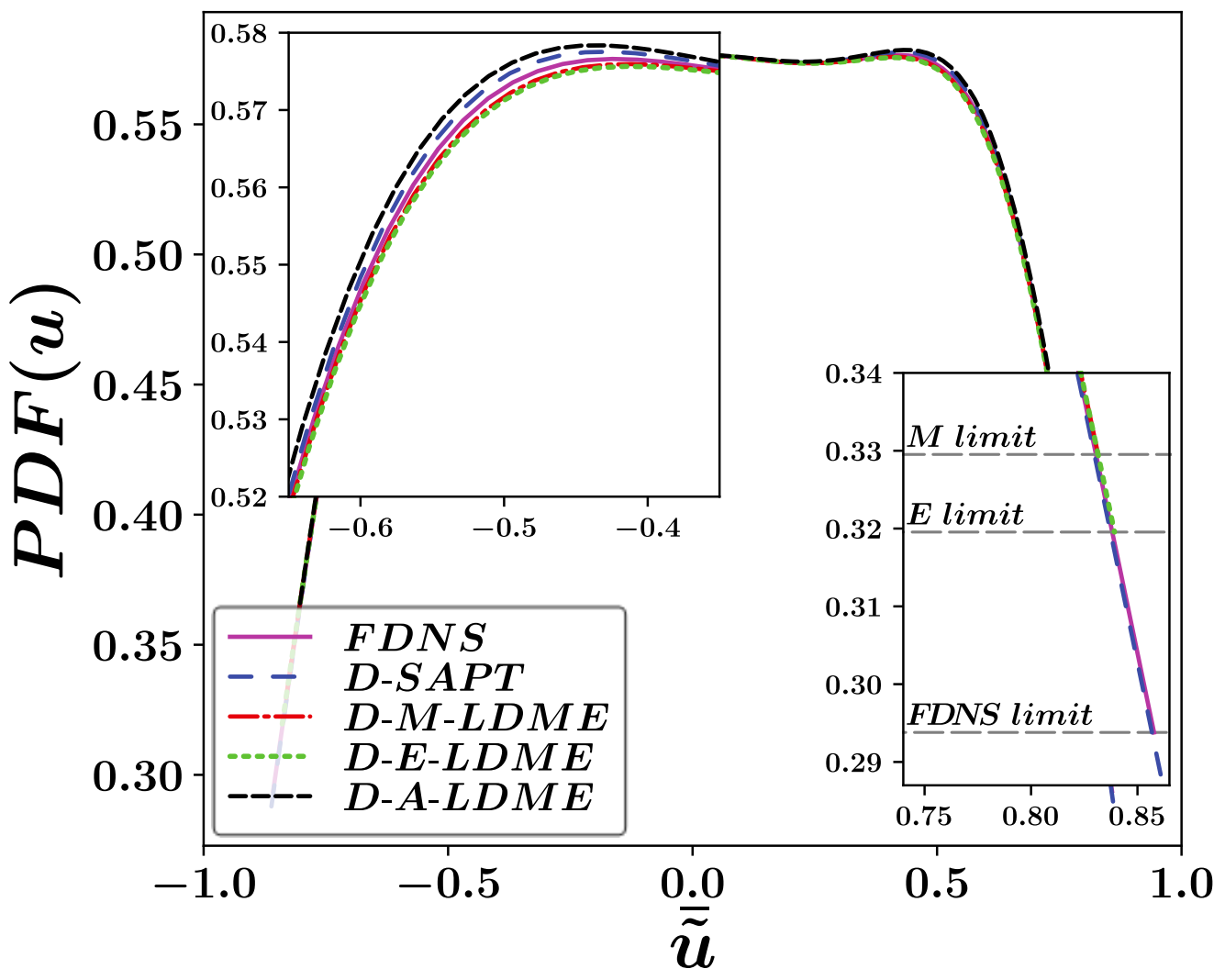

Figure 7: The "decaying" benchmark: The velocity PDF, with PT-based (SAPT) and different mixed AD-DEV closures against FDNS at $t = 2.5$. The extent of the tail of each PDF is indicated on the right inset for the sake of clarity.

*5.2. Comparison of the new and conventional closures: Burgers turbulence*

The objective of this section is to compare the new mixed AD-DEV strategies with conventional pure DEV, standard mixed AD-DEV, and PT-based closures available in the literature. In addition, we check whether the findings of the previous section are valid for all three benchmarks. In figure 8a, a comparative analysis is conducted between the D-pure-LDME as a conventional pure EV-based LES, D-SAPT as the best PT-based AD-LES, D-S-LDME as a conventional mixed AD-LES, and our novel strategies, namely D-M-LDME, D-E-LDME, and D-A-LDME. It is manifested that all AD-LESs perform better than the pure DEV-LES. In addition, the new mixed models, especially E-LDME and A-LDME, show better performance than the standard mixed model and even than the computationally intensive SAPT.

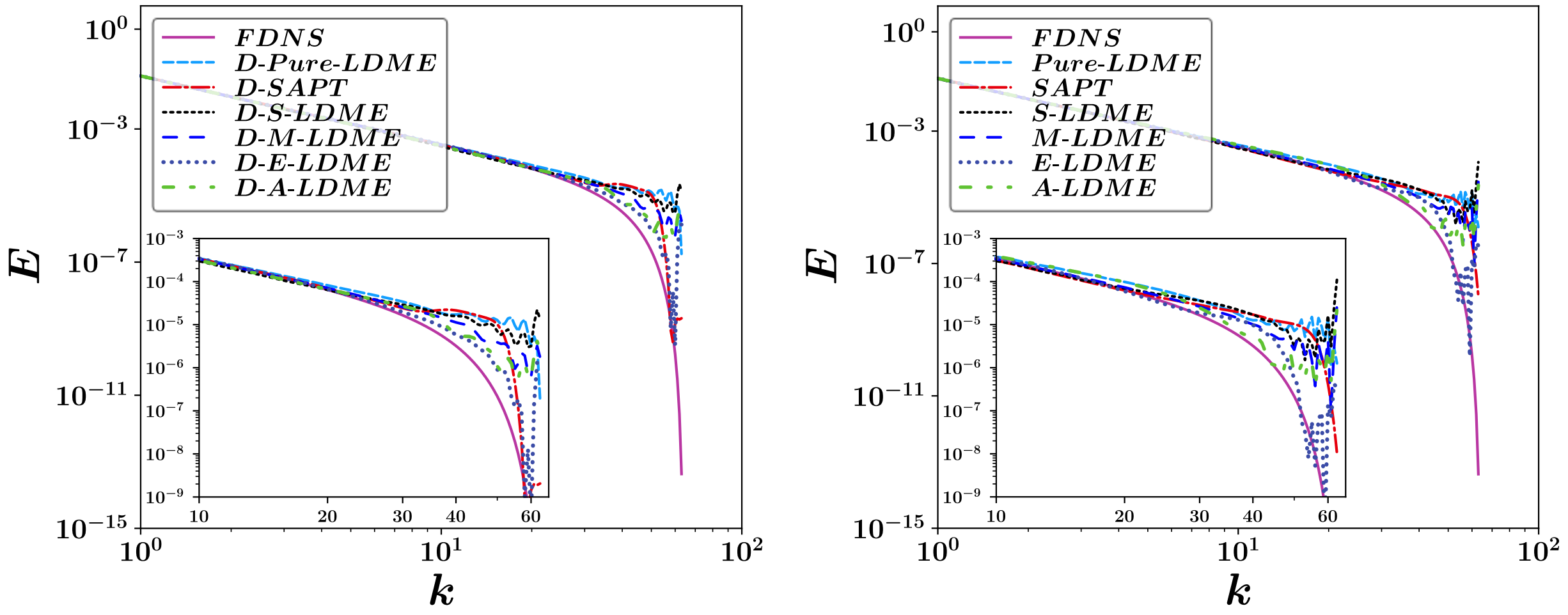

(a) (b)

Figure 8: The "decaying" benchmark: The energy spectrum, with pure DEV LES, PT-based AD-LES, and different mixed AD-LES approaches against FDNS at $t = 2.5$. a) The de-aliased and b) regular formulation.

Although the de-aliased formulation is recommended, the limitation of available solvers may enforce the use of regular formulation. The regular formulation has been used in the majority of available LES works. Therefore, the performance of the models using regular formulation is also of interest. Figure 8b provides the model comparison using the regular formulation, corresponding to the de-aliased formulation in figure 8a. Despite the existence of additional aliasing errors in the results, the conclusions made for the de-aliased formulation are still valid using the regular formulation, i.e., the improved predictions by all new mixed models compared to the conventional closures and the notable performance of E-LDME and A-LDME.

To check the validity of the findings for the other benchmarks, the comparison of different models for the injection benchmark is reported in figure 9. According to figure 9a, the models show similar characteristics as in the decaying test, and E-LDME and A-LDME are still the best choices. To check the sensitivity of the results of the mixed models to the choice of the dynamic EV closure, figure 9b illustrates the predictions of different models, this time, combined with the widely-used DSM. Although LDME is preferable to DSM (Mokhtarpoor & Heinz 2017), the trend of the predictions using DSM (figure 9b) resembles the ones using LDME (figure 9b), and it is evidenced that the novel closures provide more accurate results with other DEV models too.

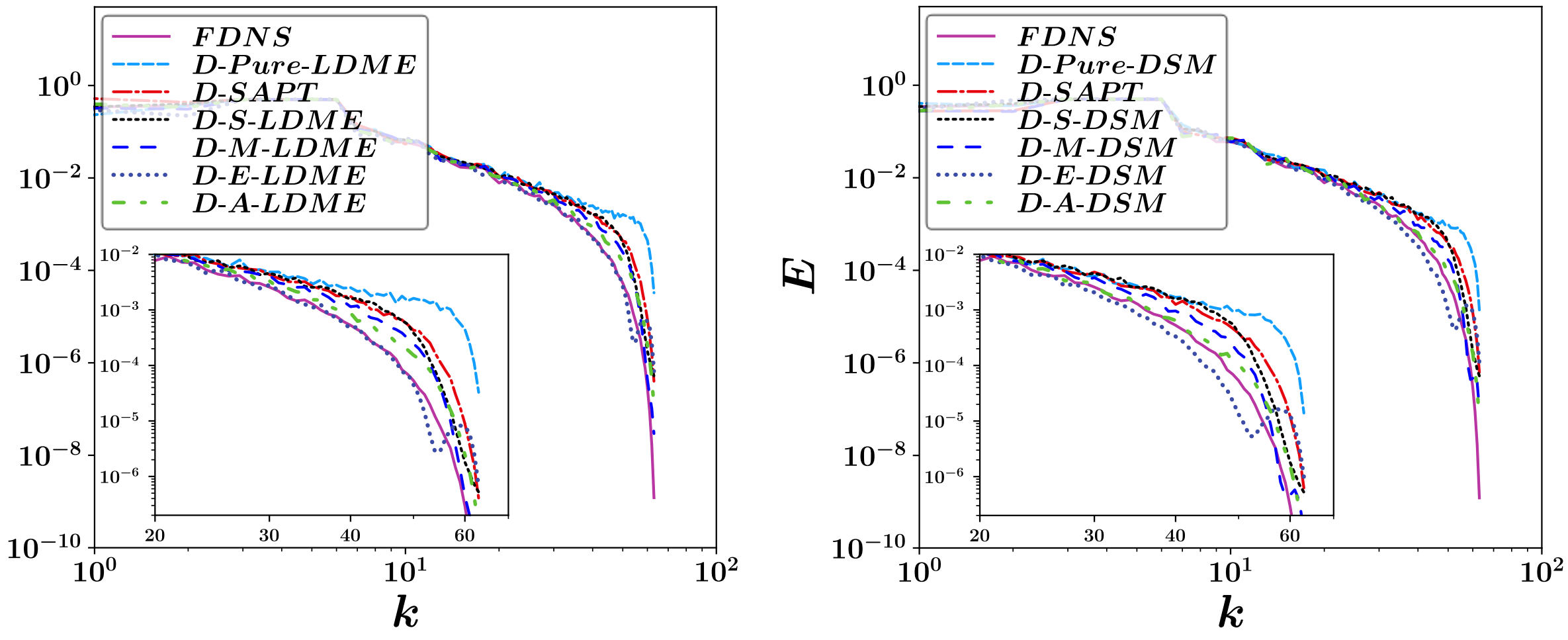

Figure 9: The "injection" benchmark: The energy spectrum, with pure DEV LES, PT-based AD-LES, and different mixed AD-LES approaches, combined with a) LDME and b) DSM closure, against FDNS.

The stochastic benchmark (Appendix C.3) is a challenging test of a highly noisy nature. Figure 10 illustrates the comparison of different models for this benchmark. As observed, the pure DEV-LES (D-Pure-DSM) is contaminated with a considerable energy pile-up, which is effectively alleviated by using AD-LES variants. However, the SAPT model suffers from an excessive dissipation and large underprediction of the energy spectrum within $\omega_C' < \omega < \omega_N'$ in this benchmark. Since no similar behavior is observed in the M-DEV and E-DEV results, it can be concluded that the observed dissipation does not originate from the deconvolved stress. This can be justified considering that unlike the decaying and injection cases, which exhibit only a few shocks in the velocity field, the stochastic case is characterized by a highly noisy nature.

This inherent noise leads to a substantial difference between the deconvolved velocity field $\overline{u^{\star}}$ and the actual velocity $\bar{\bar{u}}$. This significant difference generates an increased level of modeled SFS stress contribution by the SAPT model according to Eq. (24) which is of a purely dissipative character. Therefore, it can be inferred that the mixed AD-DEV models demonstrate superior performance in this case. Nevertheless, conventional S-DEV (both S-LDME and S-DSM) closures diverged for this case due to the instabilities induced by their inconsistencies (see table 2). On the other hand, all new mixed models perform well and are close to the FDNS result.

Finally, a quantitative error analysis based on the results reported in figure 8-figure 10 is conducted in table 5. The norm of error of the energy spectrum within the range $\omega_C' < \omega < \omega_N'$, where the results deviate noticeably from FDNS, is computed by Eq. (56) and reported in this table for all three benchmarks. Based on this data, the pure DEV-LES has the largest error in all benchmarks, advocating the superiority of AD-LES. Moreover, the standard mixed AD-DEV brings about the largest error among all AD-LES and even diverges for the stochastic benchmark. In addition, all new mixed models generally outperform the computationally intensive SAPT with a distinctly excellent performance of E-DEV, followed by A-DEV.

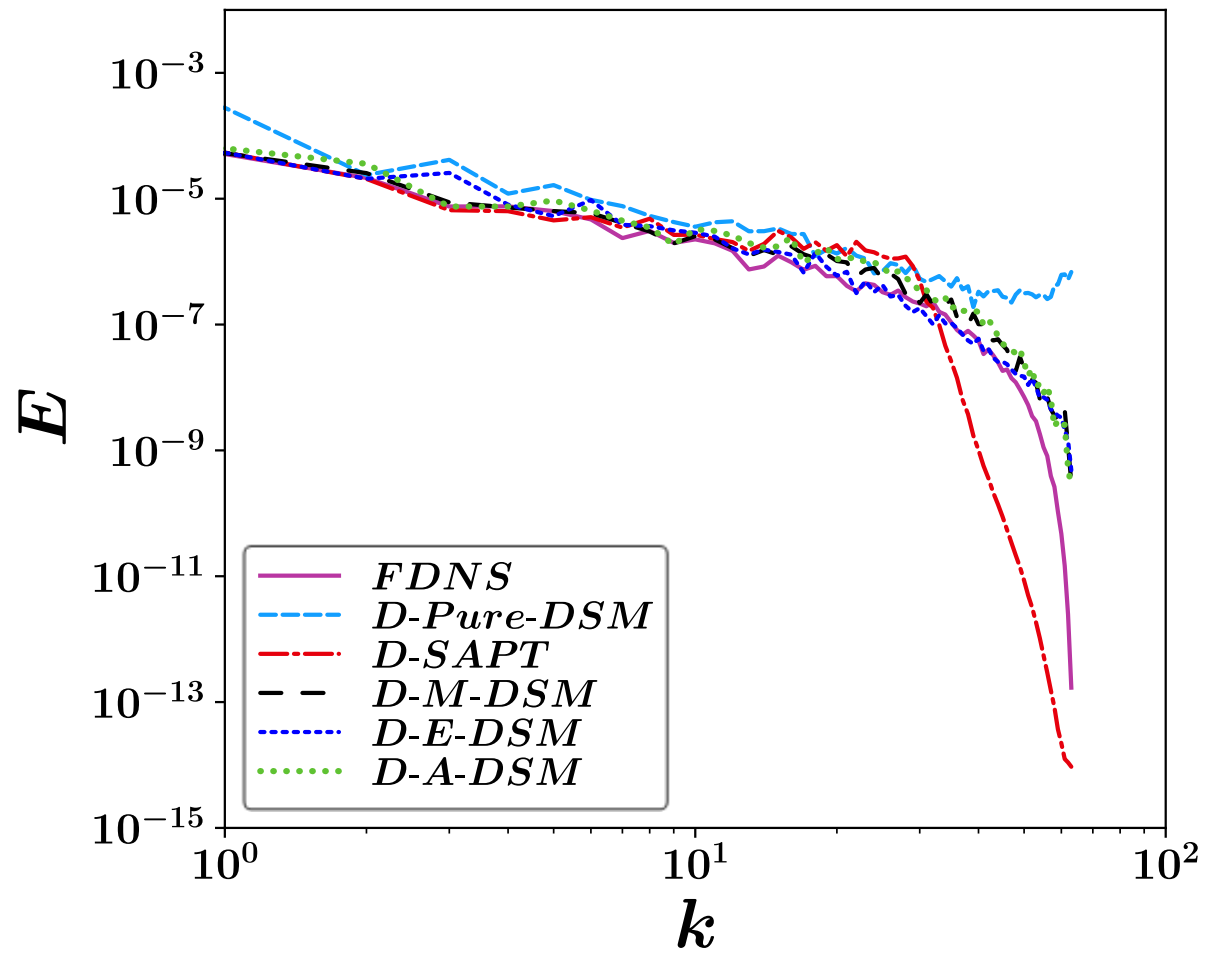

Figure 10: The "stochastic" benchmark: The energy spectrum, with pure DEV LES, PT-based AD-LES, and different mixed AD-LES approaches, combined with DSM closure, against FDNS. Note that S-DSM diverged.

Table 5: The error norm of the energy spectrum (within $k_C < k < k_N$), with pure DEV LES, PT-based AD-LES, and different mixed AD-LES approaches for all benchmarks (corresponding to figure 8-figure 10).

| Benchmark | D-Pure-DEV | D-SAPT | D-S-DEV | D-M-DEV | D-E-DEV | D-A-DEV |
|---|---|---|---|---|---|---|
| **Decaying** | 3.87 | 2.37 | 2.23 | 1.24 | 0.45 | 0.99 |
| **Injection** | 1.89 | 1.51 | 1.53 | 1.17 | 0.86 | 0.93 |
| **Stochastic** | 1.47 | 1.01 | Diverged | 1.04 | 0.79 | 1.03 |

### *5.3. The new and conventional AD-LES closures: Turbulent channel flow*

In this section, the performance of the AD-LES models in the channel flow benchmark is evaluated. Figure 11 presents the flow statistics predicted by the new and conventional AD-LES closures against the reference DNS data. According to this figure, the predictions by the new M-LDME and A-LDME models are close to the ones by the conventional S-LDME in terms of primary statistics, namely the mean velocity (figure 11a) as well as streamwise normal (figure 11b) and shear (figure 11e) Reynolds stress components. However, notable improvements are observed in the prediction of the other Reynolds stress components by M-LDME and A-LDME compared to the conventional models (figure 11c and d), with the best performance pertains to M-LDME in this case. In contrast to the Burgers benchmarks, E-LDME exhibits a lower accuracy than the other models for the turbulent channel flow, showing a greater positive log-layer mismatch in the mean velocity profile and overprediction of the streamwise normal Reynolds stress. To justify these observations, the energy cascade flux across the filter length-scale, $\Pi_C = -T_{ij}\bar{\tilde{S}}_{ij}$, whose negative values are indicative of the backscatter level, is considered (Yao *et al.* 2024). The predicted dimensionless energy cascade flux for each model at $y^+ = 18$ plane, within the inner region of the turbulent boundary layer where the production rate of the kinetic energy is dominant and $\langle u_1' u_1' \rangle$ takes a peak, is illustrated in figure 12. The largest negative values of $\Pi_C^+$ by E-, M-, A-, and S-LDME are $-0.81$, $-0.39$, $-0.36$, and $-0.28$, respectively. Figure 12 reveals that the E-LDME model shows the strongest inverse cascade, which is evident by the largest negative value of $\Pi_C^+$, compared to the other models. This suggests an excessive transfer of energy from unresolved to resolved scales by this closure, leading to the observed statistical data overprediction (Davidson 2013).

Additionally, figure 11 shows that the standard S-DEV model demonstrates a much better accuracy in predicting primary statistics, i.e., $\langle u_1 \rangle$, $\langle u_1' u_1' \rangle$, and $\langle u_1' u_2' \rangle$, in the channel flow compared to the Burgers benchmarks. To analyze this point, the deconvolved and modeled stress components predicted by different models are compared in figure 13. As it can be seen, the CI-inconsistency is still evident in S-LDME results and the modeled SFS (or SGS) stress tensor components are extremely small and close to zero, on average, using this model, see figure 13b and d. However, for this benchmark, the deconvolved SFS stress of S-LDME adjusts itself by increasing its contribution to the total SFS stress, $T_{ij} = b_{ij} + a_{ij}$, compared to the other models (see figure 13a and c), compensating for the $a_{ij}$ budget. Besides the difference in physics, the order and type of the explicit filter, and discretization schemes used in the two benchmarks, the main reason underlying the better performance of S-LDME in the channel flow test compared to the Burgers benchmarks can be associated with the difference in the incorporated LES grid resolutions. More specifically, the ratio of the LES-to-DNS cell sizes in the channel flow benchmark is $\widetilde{\Delta}/h_{DNS} \sim 4$ while for the Burgers benchmarks, this ratio was $\widetilde{\Delta}/h_{DNS} \sim 16-64$. Therefore, to examine the effect of LES grid resolution on the performance of the models, AD-LES solutions using a coarser LES grid with $\widetilde{\Delta}/h_{DNS} \sim 5.5$ and increased contribution of SGS stress are obtained. The error norms of different statistics for the default and coarse grids are computed and reported in table 6. Based on this data, the new M-LDME and A-LDME closures demonstrate their superiority, even in predicting the primary statistics, to the standard S-LDME model as the LES-to-DNS cell-size ratio grows. This is due to the increased effect of the modeled stress, $a_{ij}$, term and pronounced impact of CI-inconsistency in S-LDME when larger $\widetilde{\Delta}/h_{DNS}$ values are adopted. Even, the E-LDME model performs better than S-LDME on the coarse grid which is in accordance with the findings for the Burgers benchmarks.

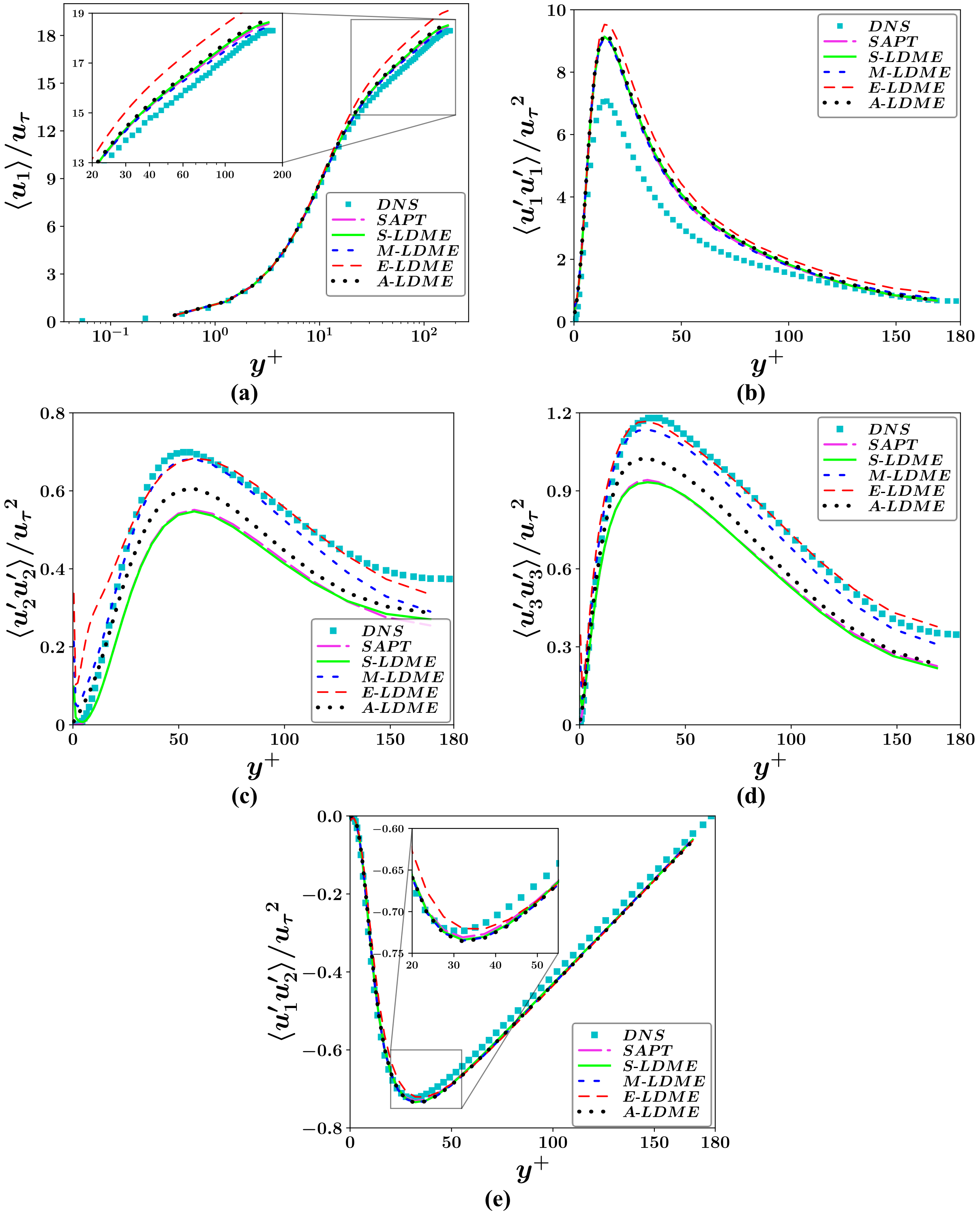

Figure 11 The “channel flow” benchmark: The comparison of mixed AD-LES models against the DNS data (Moser *et al.* 1999) for the mean streamwise velocity (a) and Reynolds stress components: streamwise (b), wall-normal (c), spanwise (d), and shear (e).

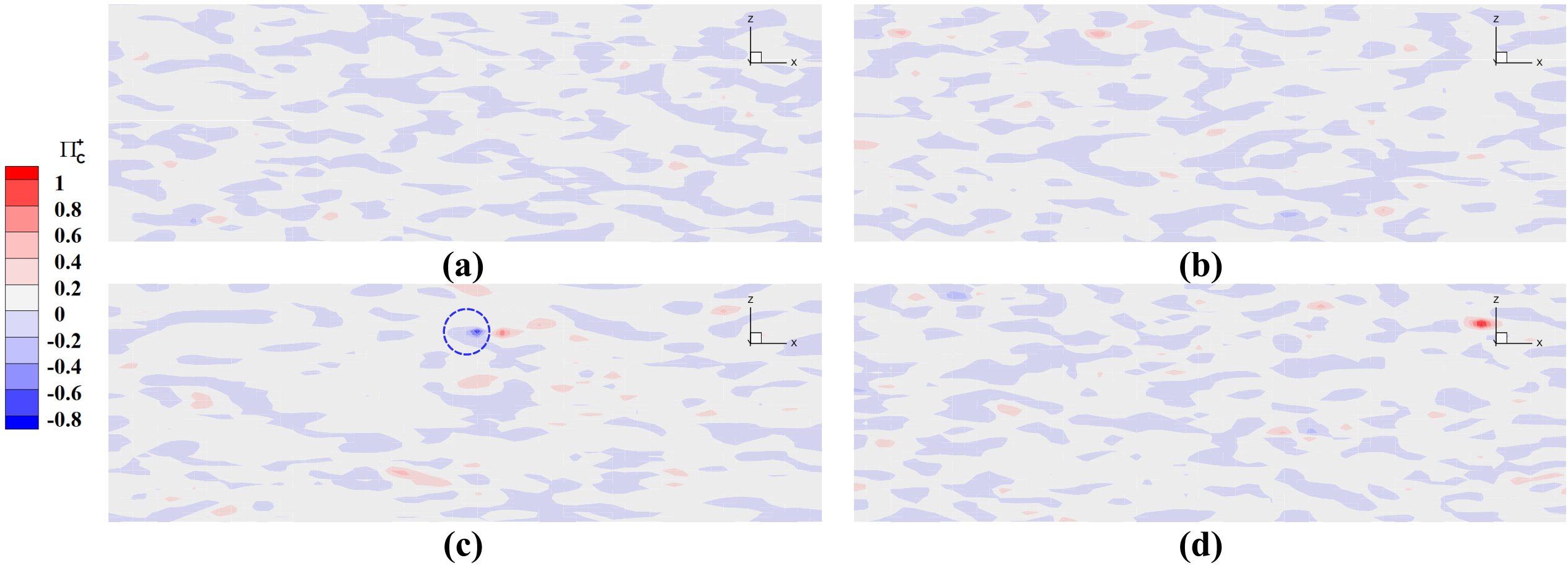

Figure 12: The "channel flow" benchmark: The instantaneous energy cascade flux, $\Pi_C^+ = \nu\Pi_C/u_\tau^4$, at the plane $y^+ = 18$, predicted by a) S-LDME, b) M-LDME, c) E-LDME, and d) A-LDME. The blue dashed circle indicates the strongest backscatter region.

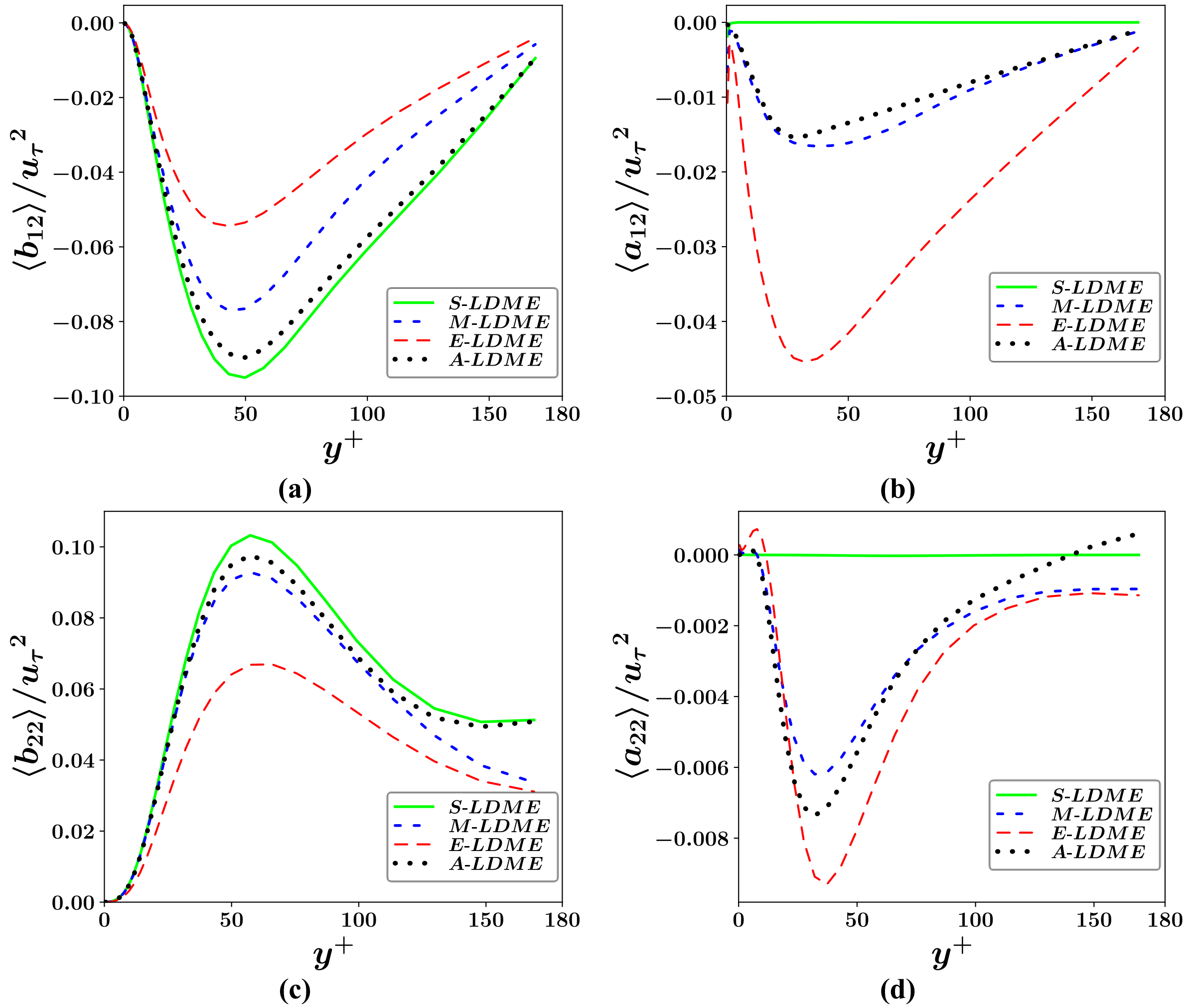

Figure 13: The "channel flow" benchmark: The comparison of the deconvolved, $b_{ij}$, and modeled, $a_{ij}$, SFS stress components profiles using different mixed AD-LES approaches.

Table 6: The "channel flow" benchmark: The error norm of the prediction of statistics by different AD-LES approaches on the two LES grid resolutions introduced in table 1.

| Grid | Statistics | Error norm | | | | |
|---|---|---|---|---|---|---|
| | | S-LDME | M-LDME | E-LDME | A-LDME | SAPT |
| Default ($\tilde{\Delta}/h_{DNS}$~4) | $\langle u_1 \rangle$ | 0.03 | 0.02 | 0.07 | 0.03 | 0.03 |
| | $\langle u_1' u_1' \rangle$ | 0.31 | 0.30 | 0.39 | 0.32 | 0.36 |
| | $\langle u_2' u_2' \rangle$ | 0.27 | 0.13 | 0.24 | 0.18 | 0.27 |
| | $\langle u_1' u_2' \rangle$ | 0.04 | 0.04 | 0.06 | 0.04 | 0.04 |
| Coarse ($\tilde{\Delta}/h_{DNS}$~5.5) | $\langle u_1 \rangle$ | 0.12 | 0.09 | 0.13 | 0.11 | 0.12 |
| | $\langle u_1' u_1' \rangle$ | 0.95 | 0.80 | 0.82 | 0.90 | 0.90 |
| | $\langle u_2' u_2' \rangle$ | 0.37 | 0.15 | 0.30 | 0.17 | 0.37 |
| | $\langle u_1' u_2' \rangle$ | 0.09 | 0.08 | 0.09 | 0.09 | 0.10 |

While M-DEV predicts the most accurate statistics for the channel flow test, A-LDME distinguishes itself from all other AD-LES models by its highest stability. It requires no clipping of the SGS eddy-viscosity, $\nu_t$, in such a manner that the results of A-LDME presented for the channel flow in this section were obtained with no *ad hoc* clipping. Conversely, the S-DEV, M-DEV, and E-DEV models required the implementation of the PTV clipping method to avoid solution divergence. Figure 14 plots the instantaneous profile of the SGS eddy-viscosity across the channel height for different mixed models. As can be seen, the viscosity ratio, $\nu_t/\nu$, in A-LDME can reach a value as small as $-30$ without instability issues while the other models are confined to the allowed lower bound of $-1$.

Finally, the new and conventional AD-LES models are evaluated in terms of the computational cost. The computational cost metrics for different models are reported in table 7 in order of their contribution to the total CPU time. The normalized CPU time is the ratio of the CPU time of each model to the one of SAPT for 30 $FTT$ simulation. The time-step size, $\Delta t$, and the number of iterations of the velocity, $N_U$, and pressure, $N_p$, equations are averaged values computed over the first 30 $FTT$ of each simulation. According to table 7, the most cost-intensive model is SAPT since its dynamic model coefficient necessitates two additional loops over the entire flow equations. This is reflected by its much larger total number of iterations of the pressure Poisson's equation, $N_p$, compared to the other models. Since $\Delta t$ and $N_p$ for the mixed models are close to each other, the difference in the computational costs of these models is predominantly connected with the total number of the filtering operations carried out in each closure. With the lowest number of filtering, A-LDME decreases the CPU time by 34% compared to SAPT and even by 3% compared to S-LDME. Therefore, A-DEV is the computationally cheapest AD-LES model as well as the most efficient model considering the accuracy of predictions in all benchmarks.

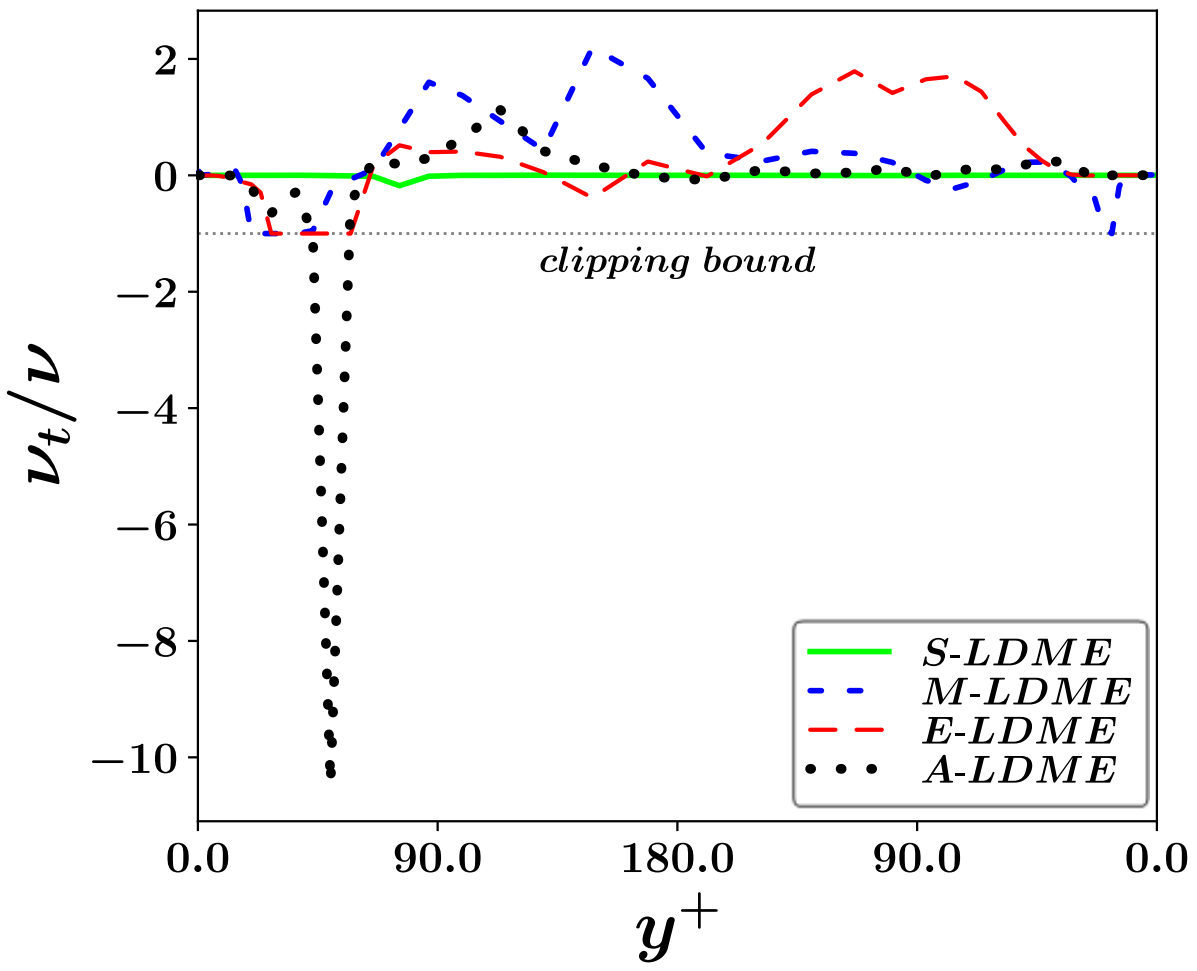

(a)

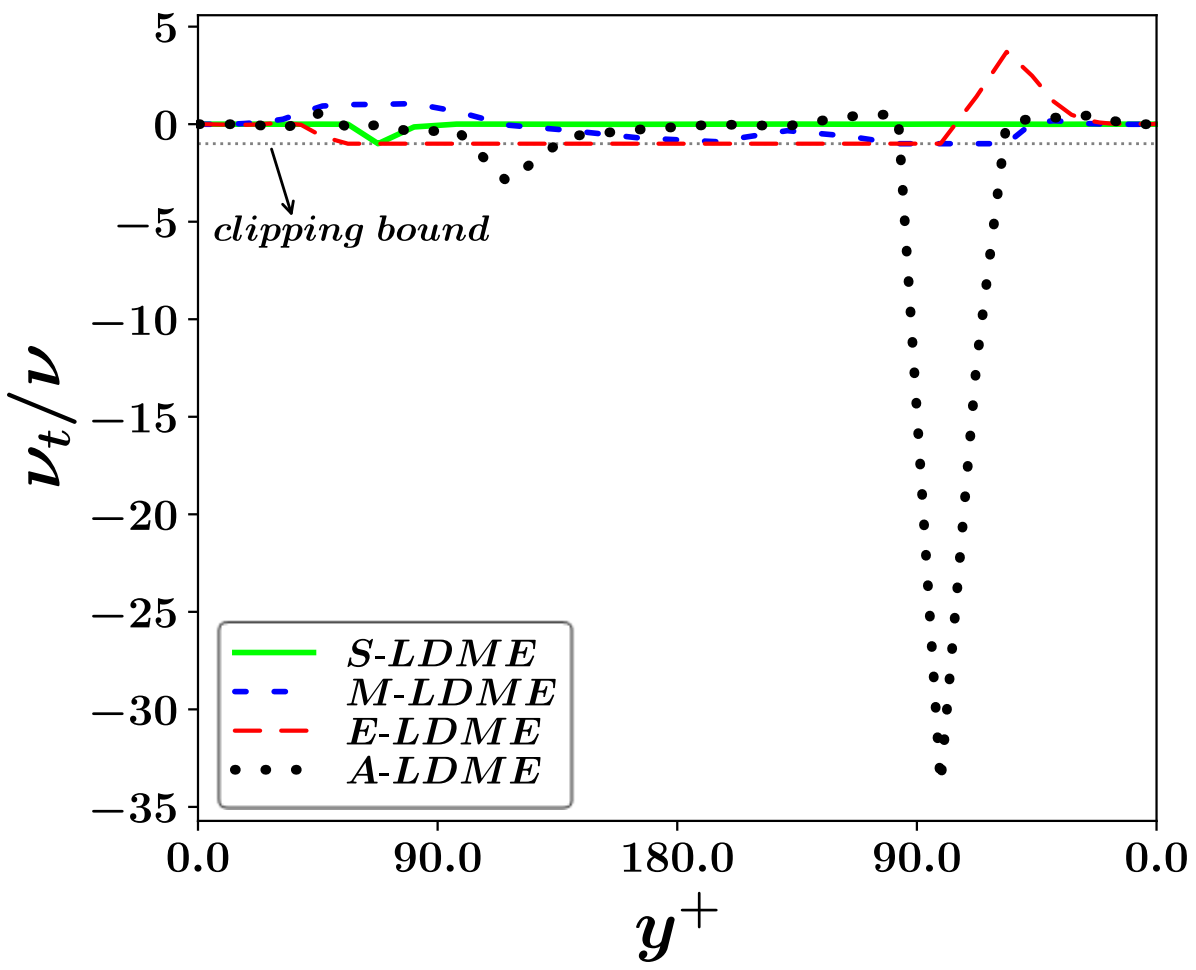

(b)

Figure 14 The "channel flow" benchmark: The comparison the instantaneous SGS eddy-viscosity, $\nu_t$, profiles using different mixed AD-LES approaches on the default (a) and coarse (b) grids.

Table 7: The "channel flow" benchmark: The computational cost metrics of different AD-LES approaches. The average number of iterations of the velocity ($N_U$) and pressure ($N_p$) equations, etc., are reported per time step.

| Metrics | Computational cost | | | | |
|---|---|---|---|---|---|
| | S-LDME | M-LDME | E-LDME | A-LDME | SAPT |
| Normalized CPU Time | 0.69 | 0.80 | 0.79 | 0.66 | 1 |
| $\Delta t \times 10^3$ (s) | 2.06 | 2.06 | 2.27 | 2.09 | 2.00 |
| $N_U$ | 2.00 | 2.00 | 2.00 | 2.00 | 6.00 |
| $N_P$ | 138.4 | 140.6 | 144.8 | 139.6 | 265.3 |
| Deconvolutions | 3 | 3 | 3 | 3 | 9 |
| Filters of width $4h$ | 18 | 24 | 27 | 0 | 1 |
| Filters of width $2h$ | 9 | 9 | 15 | 24 | 15 |
| Total filters* | 114 | 144 | 165 | 39 | 65 |

* Each deconvolution operation or filter of width $4h$ requires 5 filtering operations of width $2h$.

## 6. Conclusion

In this study, the existing mixed Standard DEV (S-DEV) and Penalty-Term (PT) hard-deconvolution models were examined and their compatibility with five consistency criteria was explored. Then, three new mixed models, namely Modified DEV (M-DEV), Enhanced DEV (E-DEV), and Alternative-DEV (A-DEV), were designed to improve the consistency characteristics of the standard model (S-DEV). After that, the performance of the aforementioned hard deconvolutions was assessed through three canonical Burgers turbulence cases and a turbulent channel flow benchmark. The results of the Burgers turbulence revealed that the novel mixed AD-DEV hard deconvolutions were successful in effectively mitigating the large unphysical oscillations near sharp velocity gradients or shocks, kinetic energy pile-up and overprediction within the filter-cut-off and Nyquist wavenumbers range, and instability problems, encountered using the standard mixed model. The superiority of all new models was proved both in the *a priori* and *a posteriori* tests. On the other hand, the turbulent channel flow test revealed that the performance of E-DEV model degrades compared to the other models as the LES grid is refined, However, M-DEV and A-DEV still demonstrated fine performance, surpassing the accuracy of the existing conventional SAPT and standard mixed S-DEV models on different grid resolutions. Finally, A-LDME was recommended as the most efficient approach which possesses the highest stability and robustness and lowest computational cost in every benchmark considered here while offering the best overall accuracy, on average for all benchmarks. The new mixed closures can make considerable progress in the prediction of turbulent flows, and their examination in more complicated engineering problems as well as developing improved models adhering to all consistency criteria introduced in this work would be interesting topics for future studies.

**Declaration of Interests:** The authors report no conflict of interest

## Supplementary Material

The supplementary material provides the comparison of the two well-known penalty-term hard deconvolutions, namely SAPT and SPT.

## Acknowledgments

This work was partially supported by TÜBITAK [grant number 221M421].

## Appendix A: Pure Dynamic Eddy-Viscosity (DEV) SGS models

In pure functional DEV approaches, Eqs. (4)-(6) are solved where SGS stress tensor, $\tau_{ij}$, is modeled by:

$$-\tau_{ij}^{M,r} = 2\nu_r \tilde{S}_{ij} = C_D m_{ij}(\tilde{u}_l); \tau_{ij}^{M,r} = \tau_{ij}^{M} - \frac{1}{3}\tau_{kk}^{M,r}\delta_{ij} \tag{A1}$$

$$m_{ij}(\tilde{u}_l) = 2\widetilde{\Delta}^2 \tilde{S}\tilde{S}_{ij} \tag{A2}$$

where the superscript $r$ shows the deviatoric part of a tensor, $\tilde{S} = (2\tilde{S}_{ij}\tilde{S}_{ij})^{1/2}$, and the model coefficient, $C_D$, is determined dynamically based on the assumptions of a chosen model. Here, two variants of DEV closures are considered. The most common model is the Dynamic Smagorinsky Model (DSM) where the model coefficient is calculated by (Germano *et al.* 1991):

$$C_D = \frac{L_{ij}^r M_{ji}}{M_{mn}M_{nm}} \tag{A3}$$

$$L_{ij} = \breve{\widetilde{u_i \tilde{u}_j}} - \breve{\tilde{u}}_i\breve{\tilde{u}}_j \tag{A4}$$

$$M_{ij} = \breve{\widetilde{m_{ij}(\tilde{u}_l)}} - m_{ij}(\breve{\tilde{u}}_l) \tag{A5}$$

where $\widetilde{(.)}$ and $\tilde{\Delta}$ represent the implicit (grid) filtering operation and its filter width which is taken as the local grid spacing ($h = \Lambda^{1/3}$ and $\Lambda$ is the local cell volume). $\breve{(.)}$ and $\breve{\Delta}$ indicate the test box filtering operation and its filter width which is chosen as $\breve{\Delta} \sim \breve{\tilde{\Delta}} = 2\tilde{\Delta}$.

Recently, Mokhtarpoor and Heinz (2017) suggested a modified DSM, called Linear Dynamic Model with Equilibrium assumption (LDME), by calculating $M_{ij}$ from:

$$M_{ij} = -m_{ij}(\breve{\tilde{u}}_l) \tag{A6}$$

They showed that this modification improves the consistency of the model and increases its stability. As pointed out by Mokhtarpoor and Heinz (2017), both DSM and LDME require clipping of the model coefficient, $C_D$, to maintain the consistency of the model and prevent instability issues. This matter is discussed in section 4.3.

**Appendix B: Standard Dynamic Eddy-Viscosity (S-DEV) AD-LES SGS models**

The formulae to calculate the dynamic coefficient, $C_D$, in Eq. (29) is derived by applying a test filter $\widehat{(.)}$ to Eq. (8) as:

$$\partial_t \hat{\bar{\bar{u}}}_i + \partial_j\left(\hat{\bar{\bar{u}}}_i \hat{\bar{\bar{u}}}_j\right) = -\partial_i \hat{\bar{\bar{p}}} + \partial_j\left(2\nu \hat{\bar{\bar{S}}}_{ij}\right) - \partial_j T_{ij}^T \tag{B1}$$

$$T_{ij}^T \equiv \widehat{\overline{\overline{u_i u_j}}} - \hat{\bar{\bar{u}}}_i \hat{\bar{\bar{u}}}_j = B_{ij} + A_{ij} \tag{B2}$$

The test filter type is the same as the explicit filter type $\overline{(.)}$ but of twice its width ($\hat{\Delta} \sim \hat{\bar{\bar{\Delta}}} = 2\bar{\Delta}$). Next, the Leonard stress is expressed as follows:

$$L_{ij} = T_{ij}^T - \hat{T}_{ij} = \widehat{\bar{\bar{u}}_i \bar{\bar{u}}_j} - \hat{\bar{\bar{u}}}_i \hat{\bar{\bar{u}}}_j \tag{B3}$$

Contracting Eq. (B3), i.e., $L_{ij}L_{ji}$, replacing $a_{ij}$ and $A_{ij}$ by the corresponding modeled tensors given by:

$$-a_{ij}^{M,r} = C_{\mathrm{D}} m_{ij}(\bar{\bar{u}}_l);\ m_{ij}(\bar{\bar{u}}_l) = 2\bar{\bar{\Delta}}^2 \bar{\bar{S}} \bar{\bar{S}}_{ij} \tag{B4}$$

$$-A_{ij}^{M,r} = C_{\mathrm{D}} m_{ij}\left(\hat{\bar{\bar{u}}}\right);\ m_{ij}\left(\hat{\bar{\bar{u}}}\right) = 2\hat{\bar{\bar{\Delta}}}^2 \hat{\bar{\bar{S}}} \hat{\bar{\bar{S}}}_{ij} \tag{B5}$$

and utilizing the method of minimizing the least-squares error, $C_{\mathrm{D}}$ is derived by:

$$C_{\mathrm{D}} = \frac{P_{ij}^r M_{ji}}{M_{mn} M_{nm}} \tag{B6}$$

$$P_{ij} = L_{ij} - \left(B_{ij}^M - \hat{b}_{ij}^M\right) \tag{B7}$$

where $b_{ij}^M$ and $B_{ij}^M$ are substituted in Eq. (B7) based on the selected soft-deconvolution decomposition, i.e., SSADM introduced in section 2.1. Comparing Eq. (B2) with Eq. (10), one can infer that $T_{ij}^T$ resembles $T_{ij}$, replacing the filter $\overline{(.)}$ with the double-filter $\widehat{\overline{(.)}}$. Recognizing Eqs. (12) and (13), the corresponding decomposition for $T_{ij}^T$ are obtained by replacing $\overline{(.)}$ with $\widehat{\overline{(.)}}$ as:

$$B_{ij} = \widehat{\bar{\bar{u}}_i \bar{\bar{u}}_j} - \hat{\bar{\bar{u}}}_i \hat{\bar{\bar{u}}}_j, A_{ij} = \hat{\tau}_{ij}; B_{ij}^M = \widehat{\overline{u_i^\star u_j^\star}} - \widehat{\overline{u_i^\star}}\, \widehat{\overline{u_j^\star}} \tag{B8}$$

Therefore, Eq. (B7) can be simplified to:

$$P_{ij} = \left(\widehat{\bar{\bar{u}}_i \bar{\bar{u}}_j} - \hat{\bar{\bar{u}}}_i \hat{\bar{\bar{u}}}_j\right) - \left(\widehat{\overline{u_i^\star u_j^\star}} - \widehat{\overline{u_i^\star}}\, \widehat{\overline{u_j^\star}}\right) \tag{B9}$$

**Appendix C: Burgers turbulence test cases**

Three different types of Burgers turbulence benchmarks are considered in this study.

*C.1 The "decaying" benchmark*

In this case, the forcing term is equal to zero ($f = 0$) and the velocity decays from the initial condition, due to the effect of the viscous dissipation in the absence of any kinetic energy production. For this case, $\nu$ is chosen as $5 \times 10^{-4}$ and the initial condition is set as $u(x, 0) = \sin(x)$. To ensure the resolution of all scales in DNS, a grid resolution of 8192 nodes and time step of $dt = 10^{-5}$ are required. For the *a posteriori* test, LESs are conducted on a uniform grid consisting of 128 nodes ($\tilde{\Delta}= h_{\text{LES}} = 64h_{\text{DNS}}$). The simulations are continued until $t = 2.5$.

*C.2 The "injection" benchmark*

In this case, $f = 0$ in Eq. (48), however, the kinetic energy injection is applied at low frequencies in Fourier space. In particular, the injection is carried out by, first, transforming $u$ to the Fourier space ($\hat{u}(k, t) = \mathcal{F}\{u(x, t)\}$ where $\mathcal{F}\{.\}$ indicates the discrete Fourier transform) in each time step and setting several low-frequency Fourier coefficients to a constant value like unity ($|\hat{u}(k_s, t)| = 1$ where $|.|$ returns the magnitude of a complex variable), and then, transforming it back to the physical space using the inverse discrete Fourier transform, $u(x, t) = \mathcal{F}^{-1}\{\hat{u}(k, t)\}$, (Moin 2010). Here, we choose $k_s$ =3, 4, 5, and 6 and $\nu = 0.02$ and the initial condition is set as $u(x, 0) = 10\sin(x)$. For DNS and LES, a grid resolution of 2048 and 128 nodes are utilized, respectively, ($\tilde{\Delta}= h_{\text{LES}} = 16h_{\text{DNS}}$). Additionally, the time step for both simulations is set to $dt = 10^{-5}$ and simulations are conducted until $t = 25$. The statistics are sampled at intervals of $dt = 10^{-3}$ and averaged over the time period of $10 < t < 25$.

*C.3 The "stochastic" benchmark*

In this case, the forcing term is a noise term (white in time but spatially correlated) defined by (Chekhlov & Yakhot 1995, Basu 2009):

$$f(x,t) = \sqrt{\frac{2D_0}{\Delta t}}\mathcal{F}^{-1}\left\{|k|^{\frac{\beta}{2}}\mathcal{F}\{\eta(x,t)\}\right\} \tag{57}$$

where $D_0$ and $\beta$ are the amplitude and spectral slope of the noise, $k$ the wavenumber, $\Delta t$ the simulation time step, and $\eta(x,t)$ is the standard Gaussian random field independently generated at each grid node and time. Here, we select $\nu = 10^{-5}$ and $\beta = -3/4$ to represent the scenario where the dynamics of the stochastic Burgers equation exhibit a complex multi-fractal behavior (Basu 2009). The simulation begins with a uniform zero initialization and concludes at $t = 200$. In order to achieve sufficient resolution across all scales in the DNS, it is necessary to employ a grid resolution consisting of 8192 nodes and a time step of $dt = 10^{-4}$. For LES, $\tilde{\Delta} = h_{\mathrm{LES}} = 64h_{\mathrm{DNS}}$ is chosen. The statistics are sampled at intervals of $dt = 0.1$ and averaged over the time interval of $100 < t < 200$.

## Appendix D: Verification and validation

To verify the accuracy of the structured-grid finite difference solver, the energy spectrum of the exact solution of the stochastic problem introduced in section C.3 is compared with the solution obtained using a spectral solver by Basu (2009). Figure 15 compares the energy spectrum computed by the current simulation to the reference solution. The two solutions are in fine agreement at small to intermediate wavenumbers but slightly differ at large wavenumbers near the Nyquist. These slight deviations arise due to the different discretization methods used for these solutions. Basu (2009) used spectral spatial discretization and second-order Adams-Bashford time discretization while this study employs the fourth-order compact spatial and fourth-order Runge-Kutta time discretization.

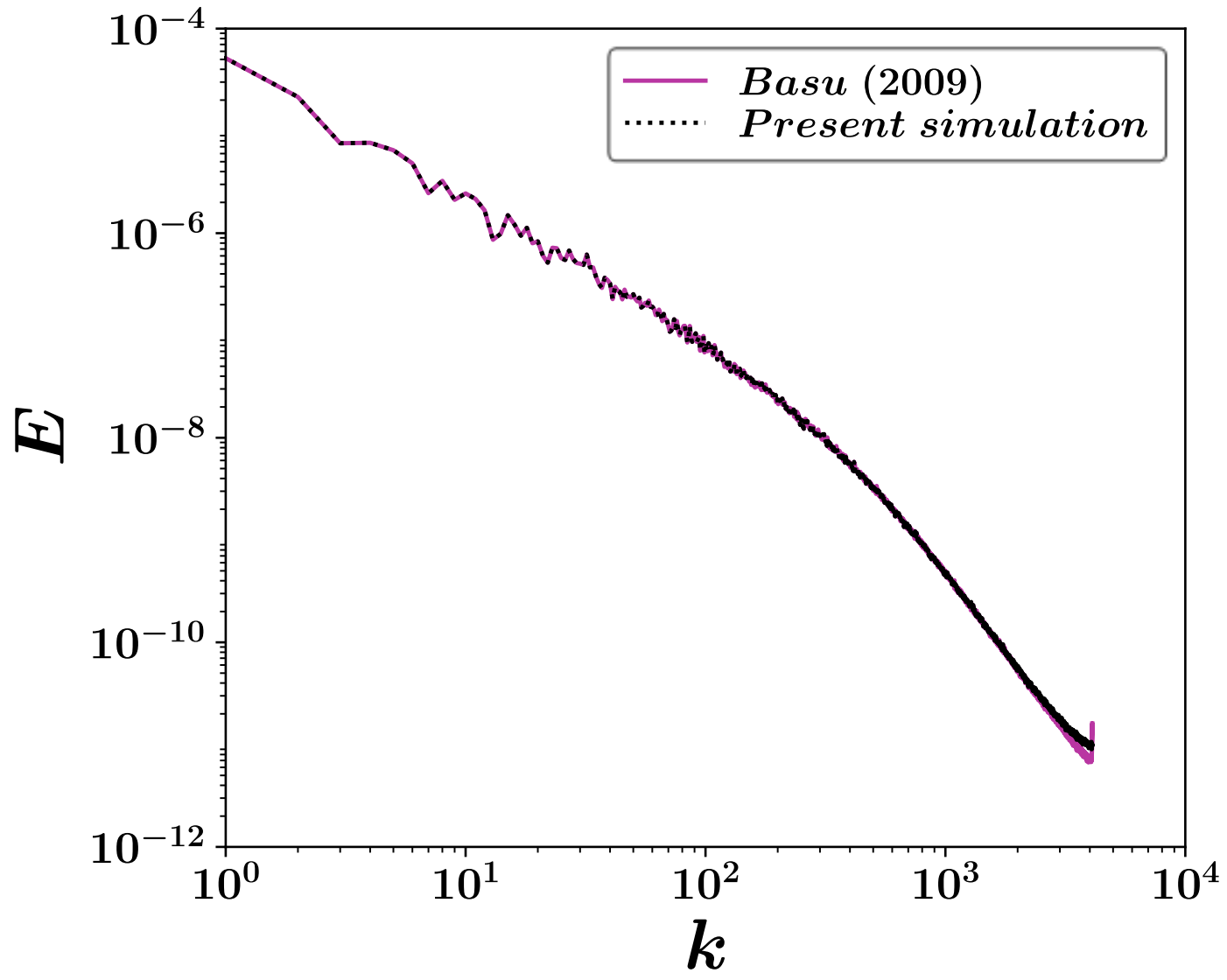

Figure 15: Verification: The (time-averaged) energy spectrum by the present numerical simulation against the reference solution (Basu 2009).

Concerning the validation of the unstructured-grid finite volume solver, i.e., OpenFOAM, the present solver has been extensively validated for LES of turbulent flows in references (Tofighian *et al.* 2020, Amani *et al.* 2023, Taghvaei & Amani 2023).